\documentclass[%
superscriptaddress,
preprint,
 amsmath,amssymb,
 aps,
]{revtex4-2}

\usepackage{graphicx,color}
\usepackage{dcolumn}
\usepackage{bm}
\usepackage[mathlines,pagewise]{lineno}

\begin{document}
\title{Quasiparticle resonance in decay spectrum of unbound nuclei near neutron drip-line}

\author{Yoshihiko Kobayashi}
\email{yoshikoba@oita-u.ac.jp}
\affiliation{Faculty of Education, Oita University, Oita 870-1192, Japan}%


\author{Masayuki Matsuo}
\email{matsuo@phys.sc.niigata-u.ac.jp}
\affiliation{Faculty of  Science, Niigata University, Niigata 950-2181, Japan}%

\date{\today}


\begin{abstract}
\begin{description}
\item[Background] The pairing correlation in weakly bound nuclei causes a mixing among bound and unbound configurations. A remarkable consequence is emergence of the quasiparticle resonance, which has been predicted with the coordinate space Hartree-Fock-Bogoliubov (HFB) theory, but not yet observed experimentally.
\item[Purpose] We discuss possible observation of quasiparticle resonances in decay spectrum of unbound nuclei near the neutron drip-line. We deal with an example of unbound nucleus ${}^{21}$C which disintegrates to ${}^{20}$C and a neutron.
\item[Method] We describe a scattering state consisting of ${}^{20}$C and a neutron in the framework of the HFB formalism. We assume that a nucleon knockout reaction produces a doorway state of the decay, and we evaluate the decay spectrum by taking an overlap of the doorway state and the scattering state of ${}^{20}\mathrm{C}+n$. A numerical calculation was performed with the Woods-Saxon potential and a density-dependent effective pairing interaction.
\item[Results] We show that the quasiparticle resonance appears as low-lying peaks in the decay spectrum of ${}^{21}$C. They originate from the weakly bound single-neutron orbits $2s_{1/2}$ and $1d_{5/2}$, but emerge as unbound resonant quasiparticle states under the influence of the neutron pairing correlation. The resonance energy and the width of the calculated quasiparticle resonances are consistent with an experimental observation whereas they are sensitive to the neutron pairing correlation.
\item[Conclusion] The results suggest that nucleon knockout reactions populating the unbound nucleus ${}^{21}\mathrm{C}$ provide realistic opportunity of experimentally observing the quasiparticle resonance and of disclosing the pairing correlation in neutron-rich nuclei.
\end{description}
\end{abstract}


\maketitle\relax\clearpage


\section{Introduction}
Exotic properties of unstable nuclei, especially neutron-rich nuclei, have been disclosed in the progress of the RI beam experiments~\cite{Tanihata2013,Nakamura2017}. The pairing correlation indispensable for the description of nuclei~\cite{BohrMottelson,RingSchuck,Dean2003,BrinkBroglia,BrogliaZelevinsky} is also affected by weak binding aspects of  unstable nuclei. The pairing correlation in such nuclei induces coupling among configurations with bound single-particle states and those involving continuum unbound single-particle states. It causes novel pairing properties such as the two-neutron halos, the di-neutron correlations, and the pairing anti-halo effects~\cite{Tanihata1985,Hansen1987,Bertsch1991,Dobaczewski1996,Meng1996,Bennaceur2000,Barranco2001,Dobaczewski2001,Myo2002,Matsuo2005,Hagino2005,Chen2014,Meng2015,Zhang2017,Hagino2017}.

Single-particle excitation mode is also influenced by the pairing correlation as is represented by the Bogoliubov quasiparticles. In finite nuclei the single-particle orbits can be either bound or unbound depending on the single-particle energy relative to the potential threshold. In pair correlated nuclei, however, even a quasiparticle state originating from a bound hole orbit may become unbound due to the coupling between the bound and unbound configurations induced by the pairing correlation~\cite{Bulgac1980,Dobaczewski1984,Belyaev1987}. A conspicuous consequence is emergence of {\it the quasiparticle resonance}, which is predicted in the Hartree-Fock-Bogoliubov (HFB) theory formulated using the coordinate space representation~\cite{Bulgac1980,Belyaev1987,Dobaczewski1996,Bennaceur1999}. However, the quasiparticle resonance has not yet been observed nor identified experimentally. It has been argued \cite{Dobaczewski1996} that observation in stable nuclei may be difficult since the quasiparticle resonance  appears at high excitation energy above the neutron/proton separation threshold of several MeV, where the level density is high.

In our preceding works~\cite{Kobayashi2016,Kobayashi2020}, we have discussed that neutron-rich nuclei having small neutron separation energy are  preferable to observe the quasiparticle resonance. In order to investigate properties of the quasiparticle resonance, we discussed the elastic scattering of neutron on a pair correlated even-even nuclei using the HFB theory. The resonance properties are analyzed in terms of the phase shift, the cross section, and the S-matrix poles of the elastic scattering. However, it is currently difficult to perform neutron scattering experiments on neutron-rich nuclei. On the contrary, breakup and/or knockout reactions of neutron-rich nuclei have been powerful tools to explore the structure of the unbound states in neutron-rich nuclei~\cite{Nakamura2017}. We note that the unbound nucleus ${}^{21}\mathrm{C}$ are studied recently with proton and neutron knockout reactions on ${}^{22}\mathrm{N}$  and ${}^{22}\mathrm{C}$~\cite{Mosby2013,Leblond2015,Orr2016}, and in a SAMURAI experiment at RIKEN RIBF low-lying peak structures are observed in the spectrum with decay channels leading to ${}^{20}\mathrm{C}+n$. Although detail analysis of the experimental results is awaited, they suggest presence of a $s$-wave peak near 1 MeV excitation and a $d$-wave peak around 1.5 MeV~\cite{Leblond2015,Orr2016} in the decay spectrum. We postulate that these peaks could be signatures of the quasiparticle resonance.

The purpose of the present study is to explore the possible quasiparticle resonance which might appear in the decay spectrum of unbound nucleus ${}^{21}\mathrm{C}$ which is produced in neutron and proton knockout reactions.  Using the same HFB theory as adopted in the preceding works~\cite{Kobayashi2016,Kobayashi2020}, we describe the decay spectrum of ${}^{21}\mathrm{C}$ disintegrating to ${}^{20}\mathrm{C}$ and a neutron. In order to make the analysis simple, we adopt an approximation adopted by Hagino and Sagawa~\cite{Hagino2016}, where the spectrum can be obtained by evaluating an overlap between a doorway state of unbound ${}^{21}\mathrm{C}$ produced in the knockout reaction, and the unbound ${}^{20}\mathrm{C}+n$ final states. We leave quantitative description of the knockout reactions with absolute cross sections for future studies.

The basic assumption of the present study is that the nuclei of our interest are under the influence of the pairing correlations. Some experimental facts which support this assumption may be remarked. In the studies of ${}^{20}\mathrm{C}$, finite occupation of neutron $2s_{1/2}$ orbit~\cite{Kobayashi2012,Togano2016} is found, suggesting an smearing of the Fermi distribution due to the pairing correlation. Suggested breaking of the $N=14$ subshell closure~\cite{Stanoiu2008,Strongman2009,Fernandez2018} in ${}^{20}\mathrm{C}$ supports also treating this nucleus as open shell.  The breaking of $N=14$ shell gap is also discussed in shell model calculations~\cite{Yuan2012,Jansen2014, Geng2022}. These studies suggest degeneracy or inversion of neutron $2s_{1/2}$ and $1d_{5/2}$ orbits. Finally the drip-line isotope ${}^{22}\mathrm{C}$ as well as ${}^{20}\mathrm{C}$ are bound nuclei while the neighbor odd-$N$ isotope ${}^{21}\mathrm{C}$ is unbound~\cite{Thoennessen2012}. It may be an indication of the neutron pairing correlation in these isotopes.

This paper is organized as follows. Section II is devoted to the theoretical framework to calculate the decay spectrum on the basis of the HFB theory. Details of the model setting and numerical calculation are discussed in Section III. We show in Section IV numerical results obtained for the neutron removal from ${}^{22}\mathrm{C}$ and then discuss the properties and the origin of the quasiparticle resonances in detail. The results for the proton removal from ${}^{22}\mathrm{N}$ are also presented. Finally, conclusions are given in Section V. In Appendix we discuss an approximation adopted in the present formulation.

\section{Theoretical Framework}
\subsection{Hartree-Fock-Bogoliubov equation in the coordinate space}
Single-particle motion under the influence of the pairing correlation is described by the Bogoliubov quasiparticle. The annihilation and creation operators, $\beta_{i}$ and $\beta^{\dagger}_{i}$, of the quasiparticle are defined by the generalized Bogoliubov transformation using the quasiparticle wave function $\phi_{i}(\bm{r}\sigma)=\left(\varphi_{1,i}(\bm{r}\sigma),~\varphi_{2,i}(\bm{r}\sigma) \right)^{T}$ in the coordinate representation and the field operators $\psi(\bm{r}\sigma)$ and $\psi^{\dagger}(\bm{r}\sigma)$~\cite{Dobaczewski1984,Matsuo2001}:
\begin{subequations}
\begin{gather}
\beta^{\dagger}_{i}=\sum_{\sigma}\int d\bm{r}\left[ \varphi_{1,i}(\bm{r}\sigma)\psi^{\dagger}(\bm{r}\sigma)+\varphi_{2,i}(\bm{r}\sigma)\psi(\bm{r}\tilde{\sigma}) \right],\\
\beta_{i}=\sum_{\sigma}\int d\bm{r}\left[ \varphi^{\ast}_{1,i}(\bm{r}\sigma)\psi(\bm{r}\sigma)+\varphi^{\ast}_{2,i}(\bm{r}\sigma)\psi^{\dagger}(\bm{r}\tilde{\sigma}) \right].
\end{gather}
\end{subequations}
The upper component of the quasiparticle wave function $\varphi_{1,i}(\bm{r}\sigma)$ may be called {\it particle component} while the lower component $\varphi_{2,i}(\bm{r}\sigma)$ is {\it hole component}. An HFB state $|\Phi\rangle$ representing the ground state of a pair correlated nucleus 
is a vacuum of the Bogoliubov quasiparticle~\cite{RingSchuck}, satisfying 
\begin{equation}
\beta_{i}|\Phi\rangle=0\quad\mathrm{for~all~}i .
\end{equation}

Imposing spherical symmetry, the radial part of the quasiparticle wave function in the partial wave expansion $\phi_{lj}(r)=\left(\varphi_{1,lj}(r),~\varphi_{2,lj}(r) \right)^{T}$ obeys the coordinate space HFB equation:
\begin{equation}
\left(
\begin{array}{ccc}
h(r)-\lambda & \Delta (r) \\
\Delta (r) &-h(r)+\lambda
\end{array}
\right)
\left(
\begin{array}{c}
\varphi_{1,lj}(r)\\
\varphi_{2,lj}(r)
\end{array}
\right)=E
\left(
\begin{array}{c}
\varphi_{1,lj}(r)\\
\varphi_{2,lj}(r)
\end{array}
\right).
\end{equation}
Here, $h(r)$, $\Delta(r)$, $\lambda$, and $E$ are single-particle Hamiltonian including centrifugal barrier, pair potential, Fermi energy, and quasiparticle energy, respectively.

The quasiparticle state becomes unbound if the absolute energy $e=E+\lambda$ is positive. In this case the particle component $\varphi_{1,lj}(r)$ is a scattering wave with asymptotic wave number $k_{1}=\sqrt{2m(E+\lambda)}/\hbar$. It represents single-particle motion of nucleon interacting with the nucleus through the mean-field $h(r)$ and the pair potential $\Delta(r)$. In the partial wave representation, the asymptotic wave function of the unbound quasiparticle state is given by
\begin{eqnarray}
&&\phi_{lj}(r,E)\nonumber\\
&&=C(E)\left(
\begin{array}{c}
\cos\delta_{lj}(E)j_{l}(k_{1}r)-\sin\delta_{lj}(E)n_{l}(k_{1}r)\\
D(E)h^{(1)}_{l}(k_{2}r)
\end{array}
\right)\nonumber\\
&&\xrightarrow[r\to\infty]{}
C(E)\left(
\begin{array}{c}
\displaystyle\frac{1}{k_{1}r}\sin\left( k_{1}r-\frac{l\pi}{2}+\delta_{lj}(E) \right)\\
0
\end{array}
\right),
\end{eqnarray}
\begin{equation}
C(E)=\sqrt{\frac{2mk_{1}}{\hbar^{2}\pi}},
\end{equation}
where $k_2$ is the wave number of the hole component $\varphi_{2,lj}(r)$~\cite{Kobayashi2016,Kobayashi2020}. The asymptotics of $\varphi_{2,lj}(r)$ is exponentially decaying  with $r$ far outside the nucleus.

In the following we denote the unbound quasiparticle state $\beta^\dagger_{lj}(E)$. Note that we neglect the magnetic substates and angular momentum algebra for simplicity.


\subsection{Decay spectrum in the Hartree-Fock-Bogoliubov formalism}
Following Ref.~\cite{Hagino2016}, we describe a decay spectrum of ${}^{21}\mathrm{C}$ with an assumption on a reaction mechanism that the knockout reaction produces a doorway state $|\Psi_{\mathrm{DWS}}\rangle$ which then decays to a scattering state of ${}^{20}\mathrm{C}+n$. With this assumption the decay spectrum may be given by
\begin{equation}
\frac{dP(e)}{de}=\left| \langle\Psi_{\mathrm{DWS}}|\Psi_{{}^{20}\mathrm{C}+n}(e)\rangle \right|^2.
\end{equation}
Here $|\Psi_{{}^{20}\mathrm{C}+n}(e)\rangle$ is the scattering state of ${}^{20}\mathrm{C}+n$ and $e$ is the kinetic energy of neutron. 

We consider the case that the daughter ${}^{20}\mathrm{C}$ is in the ground state, which we denote $|\Psi_{{}^{20}\mathrm{C}}\rangle$ in the following. We assume that it is a pair correlated ground state, and we describe it by means of the HFB framework:
\begin{eqnarray}
\left|\Psi_{{}^{20}\mathrm{C}}\right\rangle = \left|\Phi_{\nu}^{({}^{20}\mathrm{C})}\right\rangle\otimes\left|\Phi_{\pi}^{({}^{20}\mathrm{C})}\right\rangle 
\end{eqnarray}
where $\left|\Phi_{\nu}^{({}^{20}\mathrm{C})}\right\rangle$ and $\left|\Phi_{\pi}^{({}^{20}\mathrm{C})}\right\rangle $ are the HFB state vectors for neutrons and protons, respectively, obtained for ${}^{20}\mathrm{C}$.

We describe ${}^{21}\mathrm{C}$ also in the HFB framework. Since it is an odd-$N$ system, the low-lying states may correspond to one-quasineutron configurations built upon an HFB ground state. We note also that unbound ${}^{21}\mathrm{C}$ is a scattering state ${}^{20}\mathrm{C}+n$ consisting of the ground state ${}^{20}\mathrm{C}$ and an unbound neutron, which  is interacting with ${}^{20}\mathrm{C}$ through the mean-field $h^{({}^{20}\mathrm{C})}(r)$ and the pair potential $\Delta^{({}^{20}\mathrm{C})}(r)$ of ${}^{20}\mathrm{C}$ in the HFB framework. Thus we describe ${}^{21}\mathrm{C}$ as a one-quasineutron state
\begin{eqnarray}
|\Psi_{{}^{20}\mathrm{C}+n,lj}(e)\rangle=\left( \beta^{({}^{20}\mathrm{C})}_{\nu lj}(E) \right)^{\dagger}\left|\Phi_{\nu}^{({}^{20}\mathrm{C})}\right\rangle\otimes\left|\Phi_{\pi}^{({}^{20}\mathrm{C})}\right\rangle
\end{eqnarray}
where the quasineutron state, described by $\left( \beta^{({}^{20}\mathrm{C})}_{\nu lj}(E) \right)^{\dagger}$, is an unbound one which obeys the boundary condition Eq.~(4). Note that we specify the scattering states with the quasiparticle energy $E=e+|\lambda|$ and the partial wave quantum numbers $lj$. 

The decay spectrum in a specific partial wave $lj$ can be evaluated with Eq.~(6), using an overlap amplitude between the doorway state $|\Phi_{\mathrm{DWS}}\rangle$ and the scattering state $|\Psi_{{}^{20}\mathrm{C}+n,lj}(e)\rangle$. The doorway state $|\Phi_{\mathrm{DWS}}\rangle$ depends on the reactions under consideration.

Consider the case where ${}^{21}\mathrm{C}$ is produced via a neutron knockout reaction on ${}^{22}\mathrm{C}$. Here we assume ${}^{22}\mathrm{C}$ is the pair correlated ground state:
\begin{equation}
\left|\Psi_{{}^{22}\mathrm{C}}\right\rangle=\left|\Phi_\nu^{({}^{22}\mathrm{C})}\right\rangle\otimes\left|\Phi_{\pi}^{({}^{22}\mathrm{C})}\right\rangle.
\end{equation}
We suppose that the knockout reaction removes one neutron from the ground state of ${}^{22}\mathrm{C}$. The doorway state in this case may be given by 
\begin{equation}
|\Psi_{\mathrm{DWS}}\rangle=a_{1}c^{({}^{22}\mathrm{C})}_{\nu 2s_{1/2}}\left|\Psi_{{}^{22}\mathrm{C}}\right\rangle
+ a_{2}c^{({}^{22}\mathrm{C})}_{\nu 1d_{5/2}}\left|\Psi_{{}^{22}\mathrm{C}}\right\rangle+\cdots
\end{equation}
where $c^{({}^{22}\mathrm{C})}_{i}$ represents particle annihilation operator for a single-particle orbit $i$ defined in the mean-field $h^{({}^{22}\mathrm{C})}(r)$ for ${}^{22}\mathrm{C}$. Here the first two terms represent components where a neutron in the $2s_{1/2}$ and $1d_{5/2}$ orbits, the most weakly bound neutron orbits in ${}^{22}\mathrm{C}$, is removed from the ground state of ${}^{22}\mathrm{C}$. The coefficients $a_1, a_2, \cdots$ are amplitudes
for each component. 

The above doorway state, Eq.~(10), can decay by emitting a neutron in multiple decay channels with different neutron partial waves. Relevant to the spectrum probing the $J^\pi=1/2^{+}$ state of ${}^{21}\mathrm{C}$ is an overlap between the doorway state, Eq.~(10), and the scattering state in the partial wave $s_{1/2}$:
\begin{eqnarray}
&&\langle\Psi_{\mathrm{DWS}}|\Psi_{{}^{20}\mathrm{C}+n,s_{1/2}}(e)\rangle \nonumber\\
&&=a_1^* \left\langle\Psi_{{}^{22}\mathrm{C}}\right|
\left( c^{({}^{22}\mathrm{C})}_{\nu 2s_{1/2}}\right)^{\dagger}|\Psi_{{}^{20}\mathrm{C}+n,s_{1/2}}(e)\rangle \nonumber \\
&&=S_1 \left\langle\Phi_{\nu}^{({}^{22}\mathrm{C})}\right|\left(c^{({}^{22}\mathrm{C})}_{\nu 2s_{1/2}}\right)^{\dagger}\left( \beta^{({}^{20}\mathrm{C})}_{\nu s_{1/2}}(E)\right)^{\dagger}\left|\Phi_{\nu}^{({}^{20}\mathrm{C})}\right\rangle,
\end{eqnarray}
with $S_1=a_{1}^*\left\langle\Phi_{\pi}^{({}^{22}\mathrm{C})}\right|\left.\Phi_{\pi}^{({}^{20}\mathrm{C})}\right\rangle$. In the present study we do not intend to discuss the absolute value of the cross section, and hence we treat the amplitude $a_{1}$ and the coefficient $S_1$ as arbitrary energy-independent constants for simplicity.

One needs matrix element 
$\left\langle\Phi_{\nu}^{({}^{22}\mathrm{C})}\right|\left(c^{({}^{22}\mathrm{C})}_{\nu 2s_{1/2}}\right)^{\dagger}\left( \beta^{({}^{20}\mathrm{C})}_{\nu s_{1/2}}(E)\right)^{\dagger}\left|\Phi_{\nu}^{({}^{20}\mathrm{C})}\right\rangle$ between two different HFB states for neutrons, $\left|\Phi_\nu^{({}^{20}\mathrm{C})}\right\rangle$ and $\left|\Phi_\nu^{({}^{22}\mathrm{C})}\right\rangle$, corresponding to the ground states of ${}^{20}\mathrm{C}$ and ${}^{22}\mathrm{C}$, respectively. As discussed in Appendix A, this can be calculated exactly, provided that the unbound quasiparticle states are discretized, e.g. by adopting the box boundary condition. In order to evaluate matrix elements for continuum scattering quasiparticle states, however, we introduce an approximation, in which the matrix element is expressed as 
\begin{eqnarray}
&&\left\langle\Phi_{\nu}^{({}^{22}\mathrm{C})}\right|\left(c^{({}^{22}\mathrm{C})}_{\nu 2s_{1/2}}\right)^{\dagger}\left( \beta^{({}^{20}\mathrm{C})}_{\nu s_{1/2}}(E)\right)^{\dagger}\left|\Phi_{\nu}^{({}^{20}\mathrm{C})}\right\rangle \nonumber\\
&&\approx \sum_{\sigma}\int d\bm{r}\varphi^{({}^{22}\mathrm{C})}_{\nu 2s_{1/2}}\left( \bm{r}\sigma \right)\varphi^{({}^{20}\mathrm{C})}_{2,\nu s_{1/2}}\left( \bm{r}\sigma,E \right)
\end{eqnarray}
in terms of the single-particle wave function $\varphi^{({}^{22}\mathrm{C})}_{\nu 2s_{1/2}}\left( \bm{r}\sigma \right)$ of the $2s_{1/2}$ orbit in ${}^{22}\mathrm{C}$, and the hole component wave function $\varphi^{({}^{20}\mathrm{C})}_{2,\nu s_{1/2}}\left( \bm{r}\sigma,E \right)$ of the unbound quasinueutron state $\left( \beta^{({}^{20}\mathrm{C})}_{\nu s_{1/2}}(E)\right)^\dagger$ in the scattering state $|\Psi_{{}^{20}\mathrm{C}+n,s_{1/2}}(e)\rangle$. We call it {\it diagonal approximation}, and its validity is discussed in Appendix A. Thus we have
\begin{eqnarray}
&&\langle\Psi_{\mathrm{DWS}}|\Psi_{{}^{20}\mathrm{C}+n,s_{1/2}}(e)\rangle \nonumber\\
&&\approx S_1\sum_{\sigma}\int d\bm{r}\varphi^{({}^{22}\mathrm{C})}_{\nu 2s_{1/2}}\left( \bm{r}\sigma \right)\varphi^{({}^{20}\mathrm{C})}_{2,\nu s_{1/2}}\left( \bm{r}\sigma,E \right).
\end{eqnarray}

The doorway state also has a decay channel of the $d_{5/2}$ wave if the amplitude $a_2$ of the second term of r.h.s. of Eq.~(10) is not zero. In parallel to the $s$-wave case, the overlap relevant to this decay channel is given by
\begin{eqnarray}
&&\langle\Psi_{\mathrm{DWS}}|\Psi_{{}^{20}\mathrm{C}+n,d_{5/2}}(e)\rangle \nonumber\\
&&= a_2^* \left\langle\Psi_{{}^{22}\mathrm{C}}\right|
\left( c^{({}^{22}\mathrm{C})}_{\nu 1d_{5/2}}\right)^{\dagger}|\Psi_{{}^{20}\mathrm{C}+n,d_{5/2}}(e)\rangle \nonumber \\
&&=S_2 \left\langle\Phi_{\nu}^{({}^{22}\mathrm{C})}\right|\left(c^{({}^{22}\mathrm{C})}_{\nu 1d_{5/2}}\right)^{\dagger}\left( \beta^{({}^{20}\mathrm{C})}_{\nu d_{5/2}}(E)\right)^{\dagger}\left|\Phi_{\nu}^{({}^{20}\mathrm{C})}\right\rangle \nonumber \\
&&\approx S_2\sum_{\sigma}\int d\bm{r}\varphi^{({}^{22}\mathrm{C})}_{\nu 1d_{5/2}}\left( \bm{r}\sigma \right)\varphi^{({}^{20}\mathrm{C})}_{2,\nu d_{5/2}}\left( \bm{r}\sigma,E \right),
\end{eqnarray}
with $S_2=a_{2}^*\left\langle\Phi_{\pi}^{({}^{22}\mathrm{C})}\right|\left.\Phi_{\pi}^{({}^{20}\mathrm{C})}\right\rangle$. It originates from the removal of a neutron in the $1d_{5/2}$ orbit of ${}^{22}\mathrm{C}$. 

\section{Details of numerical calculation}
The HFB model adopted in the present study is based on a phenomenological Woods-Saxon mean-field and the two-body interaction of a contact type responsible for the pairing correlation.

We use a standard Woods-Saxon potential~\cite{BohrMottelson}
\begin{equation}
V(r)=V_{0}f(r)+V_{\mathrm{SO}}\left(\vec{l}\cdot\vec{s}\right)\frac{(r_{0}+\Delta r_{0})^{2}}{r}\frac{d}{dr}f(r)
\end{equation}
\begin{equation}
V_{0}=-51+33\frac{N-Z}{A}+\Delta V_{0},\quad V_{\mathrm{SO}}=22-14\frac{N-Z}{A}
\end{equation}
\begin{equation}
f(r)=\left[ 1+\exp\left( \frac{r-R}{C_{a}a} \right) \right]^{-1}
\end{equation}
\begin{equation}
a=0.67~\mathrm{fm},\quad R=r_{0}A^{1/3}~\mathrm{fm},\quad r_{0}=1.27~\mathrm{fm}
\end{equation}
while we modified slightly the parameters as represented by $\Delta V_{0}$, $C_{a}$, and $\Delta r_{0}$.  Here the radius parameter is slightly reduced by $\Delta r_{0}=-0.17$ fm so that the density distribution obtained for ${}^{20}\mathrm{C}$ reproduces the experimental root-mean-squared (r.m.s) radius $2.97^{+0.03}_{-0.05}$ fm~\cite{Togano2016}. Adjustments in the potential depth and the diffuseness, $\Delta V_{0}=-12.0$ MeV and $C_{a}=1.2$ respectively, are introduced to control the single-particle energies of the neutron $2s_{1/2}$ and $1d_{5/2}$ orbits.

We use a density-dependent delta-interaction (DDDI) for the pairing interaction~\cite{Dobaczewski2001,Dobaczewski2002a,Dobaczewski2002b}
\begin{equation}
v_{\mathrm{pair}}(r-r^{\prime})=v_{0}\left( 1-\eta\left( \frac{\rho(r)}{\rho_{0}} \right)^\alpha \right)\left( \frac{1-P_{\sigma}}{2} \right)\delta(r-r^{\prime}),
\end{equation}
where $\rho(r)=\rho_{n}(r)+\rho_{p}(r)$, $\rho_{0}=0.32~\mathrm{fm}^{-3}$, $\eta=1.0$, and $\alpha=1.0$. The above DDDI leads to the self-consistent pair potential
\begin{equation}
\Delta(r)=v_{0}\left( 1-\eta\left( \frac{\rho(r)}{\rho_{0}} \right)^\alpha \right)\tilde{\rho}(r),
\end{equation}
with $\tilde{\rho}(r)$ being the pair density which are given in terms of the quasiparticle wave functions. The pairing force strength is the most important parameter which influences the pairing correlation and the quasiparticle resonances under discussion. We will investigate dependence on the pairing correlation by controlling $v_0$. Note a typical value is $v_{0}=-292.0~\mathrm{MeV}\cdot\mathrm{fm}^{3}$~\cite{Matsuo2010}, which reproduces the experimental neutron pair gap in ${}^{120}$Sn. 

We solve the HFB equation in the coordinate space with spherical symmetry using the polar coordinate system with the partial wave expansion. We solve the radial HFB equation using the Runge-Kutta method by imposing the box boundary condition $\varphi_{i}\left( R_{\mathrm{max}} \right)=0$ with $R_{\mathrm{max}}=30~\mathrm{fm}$ and the truncation with respect to the maximal angular quantum numbers $l_{\mathrm{max}}~,j_{\mathrm{max}}=7,~(15/2)\hbar$ and the maximal quasiparticle energy $E_{\mathrm{max}}=60~\mathrm{MeV}$. We calculate continuum quasiparticle states by solving the same radial HFB equation using the pair potential and the Fermi energy thus obtained. In this case the quasiparticle wave function is connected to the asymptotic wave, Eq.~(4), at the box radius $r=R_{\mathrm{max}}=30$ fm far outside the interaction region.

\section{Quasiparticle resonance in decay spectrum}
\subsection{Decay spectrum of ${}^{21}\mathrm{C}$ in neutron knockout reaction on ${}^{22}\mathrm{C}$}
We first discuss a representative result which is obtained with the pairing force strength $v_{0}=-320.0~\mathrm{MeV}\cdot\mathrm{fm}^{3}$. Table I shows the results of the HFB calculation for the ground states of ${}^{20}\mathrm{C}$ and ${}^{22}\mathrm{C}$. Here we show the neutron Fermi energy $\lambda$, the r.m.s matter radius $\sqrt{\langle r^{2}_{m}\rangle}$, and the neutron average pair gap, estimated in two different ways
\begin{subequations}
\begin{gather}
\Delta_{uv}=\left. \int \Delta(r)\tilde{\rho}\left(\bm{r}\right)d\bm{r}\right/\int \tilde{\rho}\left(\bm{r}\right)d\bm{r},\\
\Delta_{vv}=\left. \int \Delta(r)\rho\left(\bm{r}\right)d\bm{r}\right/\int \rho\left(\bm{r}\right)d\bm{r}.
\end{gather}
\end{subequations}
The Woods-Saxon single-particle energies for the neutron $2s_{1/2}$ and $1d_{5/2}$ are also shown. Note that $1d_{3/2}$ is unbound. The two-neutron separation energy $S_{2n}$ is not calculated directly since the Woods-Saxon model is not suitable for evaluating the total energy, and instead we list here an estimate using the Fermi energy $S_{2n}\approx -2\lambda$. It is noted that the adopted force strength $v_{0}=-320.0~\mathrm{MeV}\cdot\mathrm{fm}^{3}$ gives a pairing gap about $\Delta \approx 2.3 - 3.0$ MeV for both ${}^{20}\mathrm{C}$ and ${}^{22}\mathrm{C}$, which is close to the systematic value $\Delta_{\mathrm{syst}}=12/\sqrt{A}=2.68$ and $2.56$ MeV, respectively. Note also the neutron Fermi energy $\lambda=-0.112$ MeV in ${}^{22}\mathrm{C}$ is negative and very small, indicating that this nucleus is bound only very weakly. It is qualitatively consistent with the experimentally known very small two-neutron separation energy $S_{2n}({}^{22}\mathrm{C})=-0.14\pm0.46$~\cite{Gaudefroy2012}. On the other hand, the neutron Fermi energy in ${}^{20}\mathrm{C}$ and $\lambda=-1.098$ MeV, pointing to small but relatively larger bounding energy than ${}^{22}\mathrm{C}$. We remark that there is no bound quasiparticle states in the HFB calculation for ${}^{20}\mathrm{C}$ and ${}^{22}\mathrm{C}$ with $v_{0}=-320.0~\mathrm{MeV}\cdot\mathrm{fm}^{3}$, pointing to that ${}^{21}\mathrm{C}$ whose ground state would be a one-quasineutron configuration is unbound in the present model.

\begin{table}[ht]
\caption{Ground state properties of  ${}^{20}\mathrm{C}$ and ${}^{22}\mathrm{C}$ obtained in the present calculation, listing the neutron single-particle energies $e_{2s_{1/2}}$ and $e_{1d_{5/2}}$ of the Woods-Saxon potential, the neutron Fermi energy $\lambda$, the matter r.m.s. radius $\sqrt{\left\langle r^{2}_{\mathrm{m}}\right\rangle}$, the average neutron pairing gap $\Delta_{uv}$ and $\Delta_{uu}$. The two-neutron separation energy $S_{2n}$ is an estimate $-2\lambda$. Experimental values are also shown for $\sqrt{\left\langle r^{2}_{\mathrm{m}}\right\rangle}$ and $S_{2n}$.}
\begin{ruledtabular}
\begin{tabular}{cccc}
${}^{20}\mathrm{C}$ & & Calc. & Exp. \\ \hline
$e_{2s_{1/2}}$ [MeV] & & $-2.468$ & $-$ \\
$e_{1d_{5/2}}$ [MeV] & & $-1.802$ & $-$ \\
$\lambda$ [MeV] & & $-1.098$ & $-$ \\
$S_{2n}$ [MeV] & & $2.195$ & $3.56\pm0.23$~\cite{NNDC}\\
$\sqrt{\left\langle r^{2}_{\mathrm{m}}\right\rangle}$ [fm] & & $3.019$ & $2.97^{+0.03}_{-0.05}$~\cite{Togano2016} \\
$\Delta_{uv}$ [MeV] && $2.569$ & $-$ \\
$\Delta_{uu}$ [MeV] && $3.051$ & $-$ \\ \hline
${}^{22}\mathrm{C}$ & & Calc. & Exp. \\ \hline
$e_{2s_{1/2}}$ [MeV] & & $-2.582$ & $-$\\
$e_{1d_{5/2}}$ [MeV] & & $-2.065$ & $-$\\
$\lambda$ [MeV] & & $-0.112$ & $-$\\
$S_{2n}$ [MeV] & & $0.224$ & $-0.14\pm0.46$~\cite{Gaudefroy2012}\\
$\sqrt{\left\langle r^{2}_{\mathrm{m}}\right\rangle}$ [fm] & & 3.184 & $3.44\pm0.08$~\cite{Togano2016}\\
$\Delta_{uv}$ [MeV] & & 2.327 & $-$\\
$\Delta_{uu}$ [MeV] & & 2.860 & $-$\\
\end{tabular}\end{ruledtabular}
\end{table}

Under the above condition, we calculated the decay spectrum of ${}^{21}\mathrm{C} \to {}^{20}\mathrm{C} + n $ in the neutron knockout reaction on ${}^{22}\mathrm{C}$, i.e. by using Eqs.~(6)~(13) and (14). The results are shown in Fig.~1, where the red solid curve is a plot for the $s_{1/2}$-wave decay after neutron removal from $2s_{1/2}$ orbit, and the green dotted curve is the one for the $d_{5/2}$-wave decay after removal from $1d_{5/2}$. The vertical axis is in an arbitrary unit as we put $S_1=S_2=1$.

\begin{figure}[ht]
\begin{center}
\includegraphics[width=86mm]{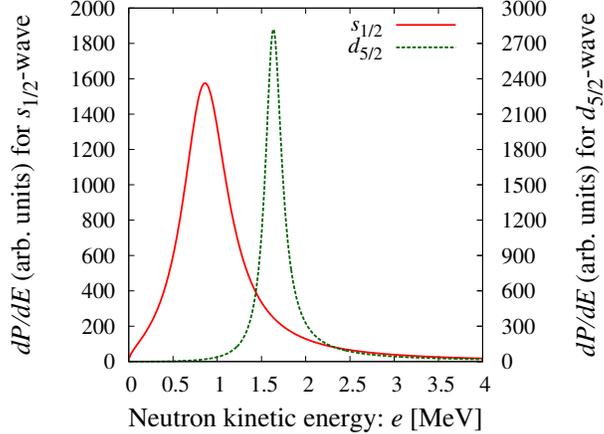}
\caption{The calculated decay spectrum $dP(e)/de$ of the ${}^{21}\mathrm{C}$ doorway state produced by the neutron removal from ${}^{22}\mathrm{C}$. The red solid curve is for decay channel ${}^{20}\mathrm{C}+n$ in the $s_{1/2}$-wave while the green dotted curve is for that in the $d_{5/2}$-wave. The absolute value is arbitrary (see text), and the horizontal axis is the neutron kinetic energy $e$.}
\end{center}
\label{fig1}
\end{figure}

A remarkable feature is that there exist peak structures indicating resonances in both $s_{1/2}$- and $d_{5/2}$-wave decays.  Note that the resonance energy $e_{R}$ and resonance width $\Gamma$, which we evaluate using the S-matrix pole~\cite{Kobayashi2020}, are consistent with the peak energy and the peak width of the experimentally observed two-peak structure in the neutron knockout reaction on ${}^{22}\mathrm{C}$~\cite{Leblond2015,Orr2016}, as shown in Table II.

\begin{table}[ht]
\caption{Resonance energy $e_{R}$ and resonance width $\Gamma$ of the quasiparticle resonances in unbound $J^\pi=1/2^{+}$ and $5/2^{+}$ states of ${}^{21}\mathrm{C}$, evaluated from the S-matrix pole for the scattering states ${}^{20}\mathrm{C}+n$ in the $s_{1/2}$- and $d_{5/2}$-waves. They are compared with energy and width of the peaks found in the experimental spectrum of the neutron knockout reaction on ${}^{22}\mathrm{C}$~\cite{Leblond2015}.}
\begin{ruledtabular}
\begin{tabular}{ccccc}
 & & Calc. & & Exp.~\cite{Leblond2015}\\ \hline
$e_{R,s_{1/2}}$ [MeV] && $0.846$ && $0.8\pm0.15$\\
$\Gamma_{s_{1/2}}$ [MeV] && $0.628$ && $0.9\pm0.9$\\
$e_{R,d_{5/2}}$ [MeV] && $1.629$ && $1.5\pm0.1$\\
$\Gamma_{d_{5/2}}$ [MeV] && $0.227$ && $0.2^{+0.9}_{-0.2}$\\
\end{tabular}
\end{ruledtabular}
\end{table}

They are the quasiparticle resonances which emerge in the HFB description of the pair correlated nuclei. We emphasize that these resonances cannot be explained as ordinary single-particle resonances. The neutron $d_{5/2}$ wave in the Wood-Saxon potential has a bound $1d_{5/2}$ orbit at $e=-1.80$ MeV. The $s_{1/2}$-wave has also a bound orbit $2s_{1/2}$ at $e=-2.47$ MeV, and hence there is neither resonance nor virtual state without taking into account the pairing.

Figure 2 shows wave functions of continuum quasiparticle states for the $s_{1/2}$ and $d_{5/2}$ resonances, evaluated at  the resonance energy in Table II. These wave functions have large amplitude inside the nucleus, indicating a clear resonance behavior. In particular, the amplitude of the hole component is larger than that of the particle component. This is a feature of {\it hole-like quasiparticle resonance} which originats from a bound hole orbit~\cite{Kobayashi2016}.

\begin{figure}[ht]
\begin{center}
\includegraphics[width=80mm]{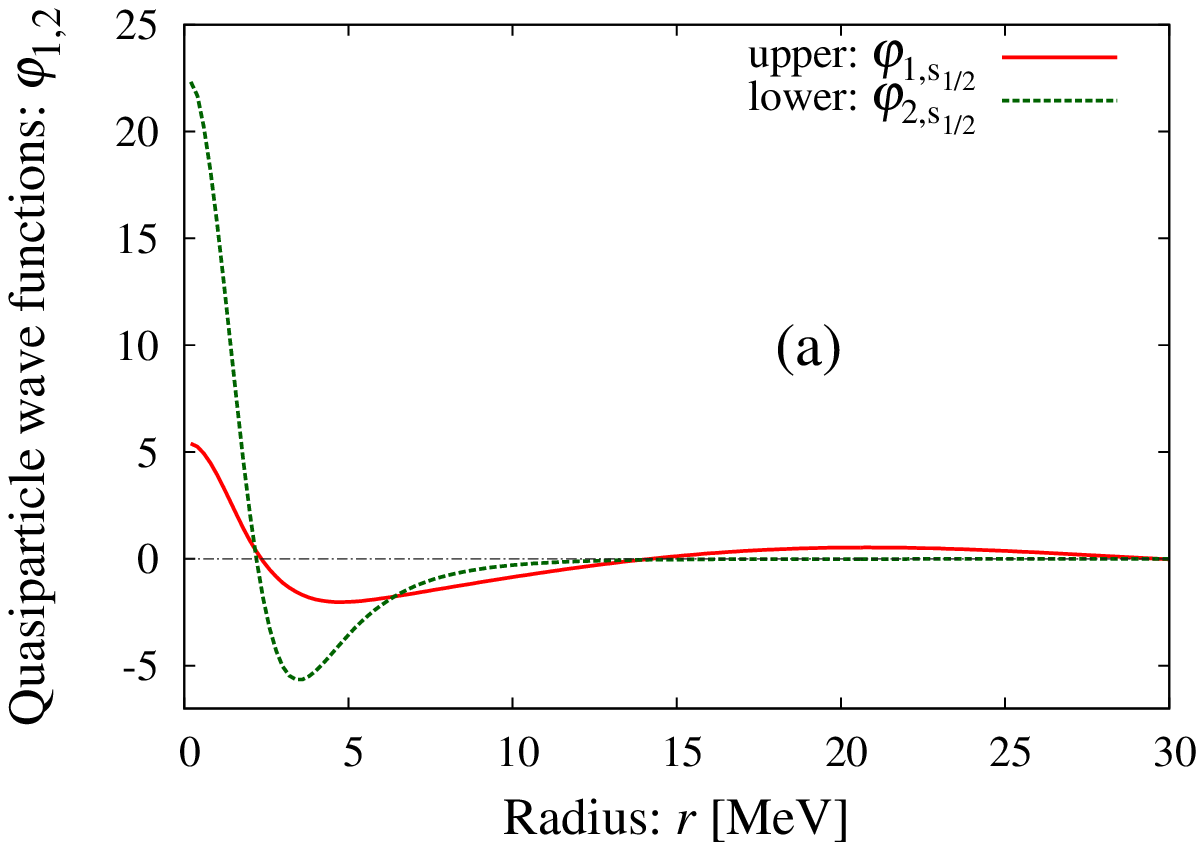}
\includegraphics[width=80mm]{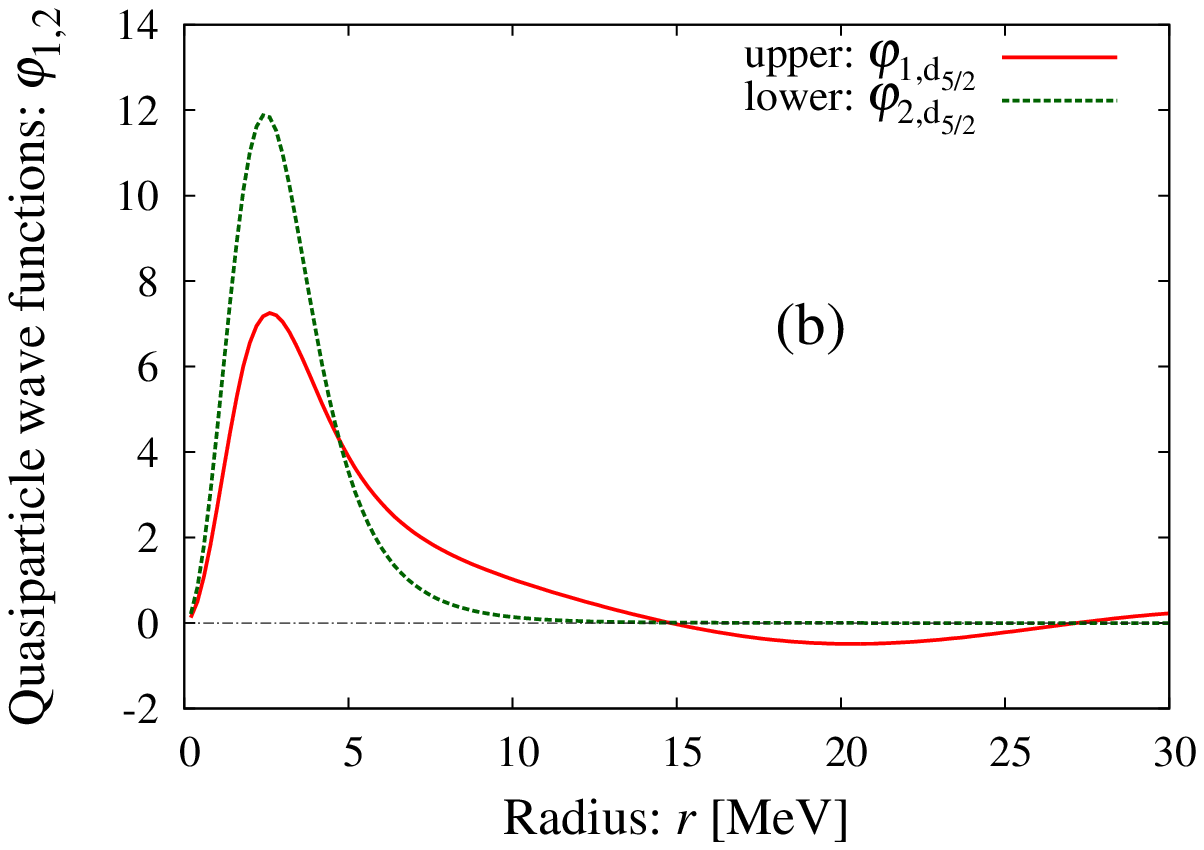}
\caption{Quasiparticle wave functions of the scattering states  ${}^{20}\mathrm{C}+n$ (a) in the $s_{1/2}$-wave and (b) in the $d_{5/2}$-wave, calculated at the resonance energies listed in Table II. The red solid and green dotted curves represent the particle and the hole components, $\varphi_{1}(r)$ and $\varphi_{2}(r)$, respectively.}
\end{center}
\end{figure}

\subsection{Dependence on pairing strength}
In order to clarify the origin and properties of the quasiparticle resonances, we examine dependence on the pairing strength. Here we vary the pairing force strength $v_0$ while the other parameters are fixed.

Figures 3 and 4 shows the decay spectrum obtained with various values of the pairing force strength $v_0$ for the $s_{1/2}$-wave decay and the $d_{5/2}$-wave decay. It is seen that the quasiparticle resonances are strongly influenced by the pairing force strength with respect to both the resonance energy and the resonance width. The resonance energy and the width increase monotonically with increasing $v_0$.

\begin{figure}[ht]
\begin{center}
\includegraphics[width=80mm]{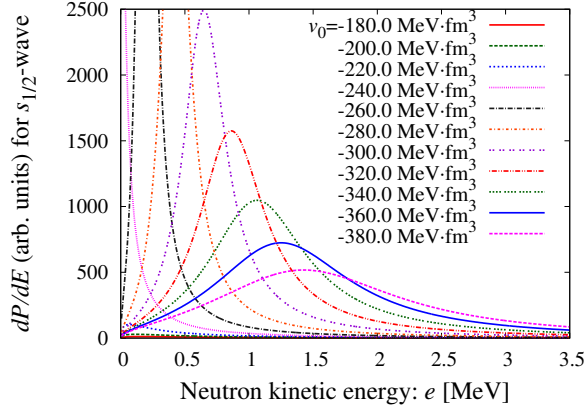}
\caption{Dependence of the decay spectrum $dP(e)/de$ on the pairing force strength $v_0$ for the ${}^{21}\mathrm{C}$ doorway state produced by the neutron removal from ${}^{22}\mathrm{C}$ for the decay channel in the $s_{1/2}$-wave.}
\end{center}
\end{figure}

\begin{figure}[ht]
\begin{center}
\includegraphics[width=80mm]{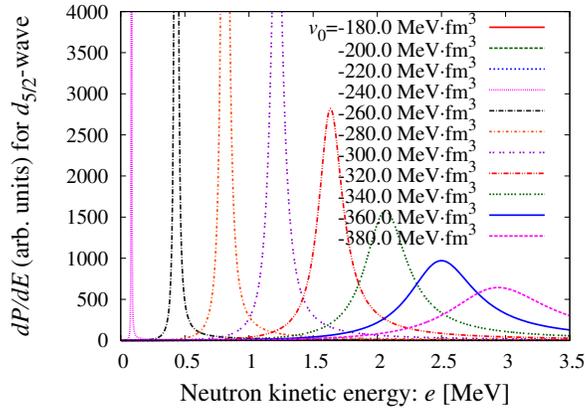}
\caption{The same as Fig. 3, but for the decay channel in the $d_{5/2}$-wave.}
\end{center}
\end{figure}

For small value of the pairing interaction strength $|v_{0}|  \lesssim 240.0~\mathrm{MeV}\cdot\mathrm{fm}^{3}$, the resonance peaks do not show up. This is because the quasineutron states associated with the $2s_{1/2}$ and $1d_{5/2}$ single-particle orbits are bound, i.e. the quasiparticle energy is below the threshold $E<|\lambda|$. Note that for $v_0=0$ the quasiparticle states are identical to the $2s_{1/2}$ and $1d_{5/2}$ single-particle orbits, which are bound at $e_{2s_{1/2}}=-2.47$ MeV and $e_{1d_{5/2}}=-1.80$ MeV. As the pairing force strength $v_0$ increases, the pair gap increases. Accordingly the quasiparticle energies of the $2s_{1/2}$ and $1d_{5/2}$ orbits increases as shown in Fig.~5. They become unbound, $E>|\lambda|$, for $|v_{0}|\gtrsim 240.0~\mathrm{MeV}\cdot\mathrm{fm}^{3}$, and emerge as the quasiparticle resonances shown in Figs.~3 and 4.

\begin{figure}[ht]
\begin{center}
\includegraphics[width=80mm]{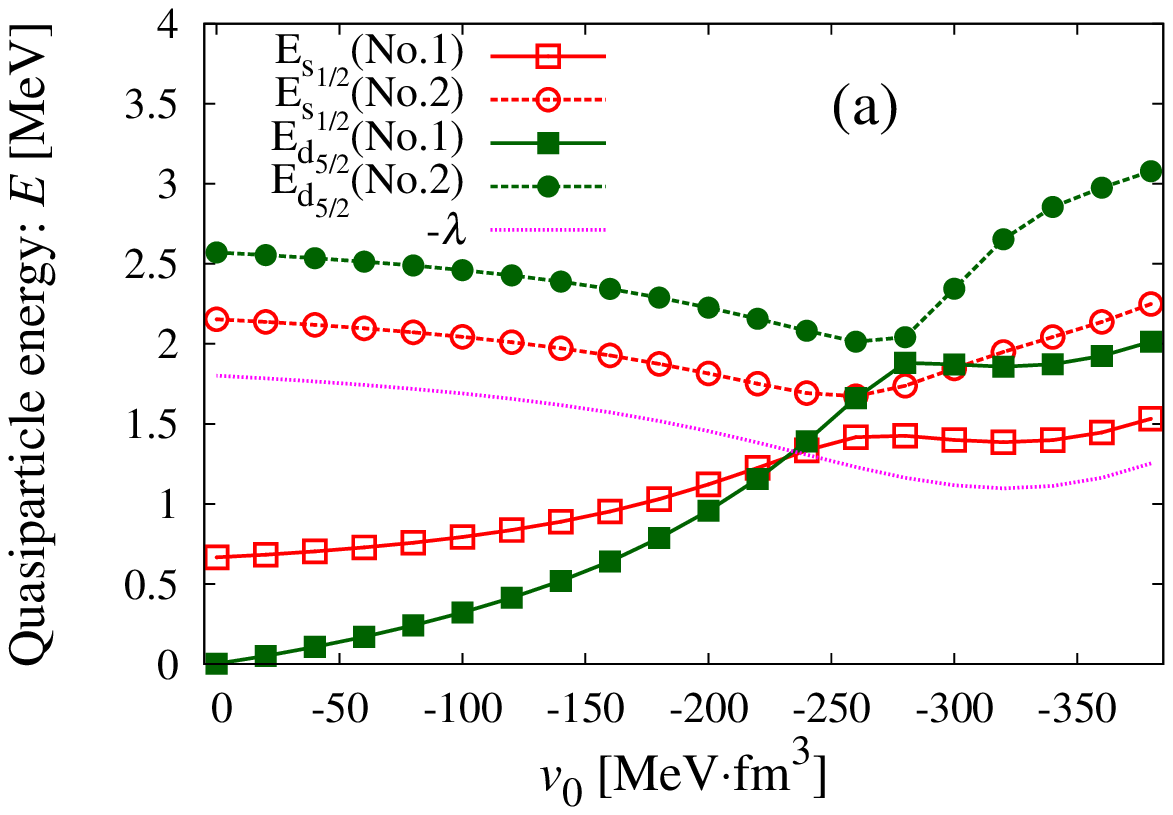}
\includegraphics[width=80mm]{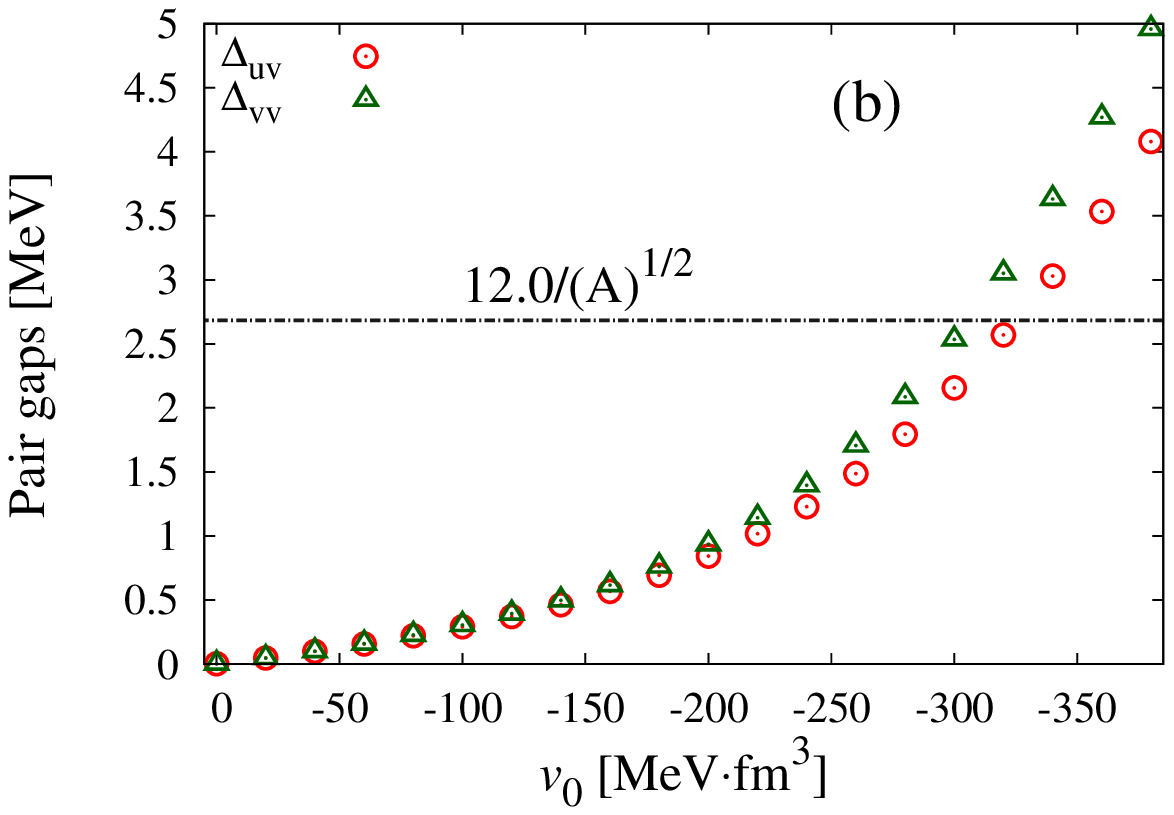}
\caption{Dependence on the pairing force strength $v_0$ of the results of the HFB calculation for ${}^{20}\mathrm{C}$. (a) The quasiparticle energies of the low-lying quasineutron states in the $s_{1/2}$ and $d_{5/2}$ waves. The threshold energy $-\lambda$ for unbound states is also shown. (b) The average neutron pair gap $\Delta_{uv}$ and $\Delta_{vv}$.}
\end{center}
\end{figure}

This behavior is very different from that of the potential scattering of a neutron, in particular in the case of the $s$-wave. In the potential scattering case, a transition from a bound state to unbound state accompanies a virtual state, but not a resonance, for the $s$-wave scattering having no centrifugal barrier. The essential difference is that the quasiparticle state in the HFB framework has a two-component wave function. Even when the transition from bound to unbound states occurs in the particle component, the hole component is confined inside the nucleus irrespective of whether the quasiparticle state is bound or unbound, see Fig.~2. Because of this property, the unbound quasiparticle state behaves like a quasi-bound state. Note also that the resonance width increases with the pairing strength (Figs.~3 and 4) since the coupling between the particle and hole components is proportional to the pairing gap.

The above behavior can be described more precisely by examining a trajectory of the S-matrix pole, the phase shift and the cross section of the neutron elastic scattering~\cite{Kobayashi2020}, which are shown in Figs.~6 and 7 for the $s_{1/2}$-wave scattering of ${}^{20}\mathrm{C}+n$. We remark here that the S-matrix is specified by two complex wave numbers $k_1$ and $k_2$ for the particle and hole components, respectively, of the quasiparticle wave functions. Correspondingly the quasiparticle energy $E=\displaystyle\frac{\hbar^2k_1^2}{2m}-\lambda=-\displaystyle\frac{\hbar^2k_2^2}{2m}+\lambda$ has four Riemann sheets. Because of the two-component structure, there appears a pair of poles even in the case  the quasiparticle state is bound: one with $\mathrm{Im}\left(k_1 \right)>0$ and $\mathrm{Im}\left(k_2 \right)>0$ corresponding to the physical bound state (located on the first Riemann sheet $E^{(1)}$, labeled '$a$' in Fig.~6) and the other one with $\mathrm{Im}\left(k_1 \right)<0$ and $\mathrm{Im}\left(k_2 \right)>0$ (on the second Riemann sheet $E^{(2)}$ with label '$b$' in Fig.~6) which exhibits a behavior of anti-bound state with $\mathrm{Im}\left(k_1 \right)<0$ while the hole component keeps a bound state character with $\mathrm{Im}\left(k_2 \right)>0$. As the quasiparticle state becomes unbound with increasing the pairing strength and pairing gap, the two poles interact on the second Riemann sheet $E^{(2)}$ and form a pair of resonance and anti-resonance poles, which represent the quasiparticle resonances discussed above. The resonance energy $e_{R}$ and the resonance width are related to the position of the pole $a$ as
\begin{equation}
e_{R}=\mathrm{Re}(E_{a})-|\lambda|,~\Gamma=-2\mathrm{Im}(E_{a}).
\end{equation}
Note that the resonance width is relatively small compared with the resonance energy. Note also that in the case of the $s$-wave quasiparticle, a virtual state is formed in a short interval just before the formation of the resonance poles~\cite{Kobayashi2020}. Influence of the virtual state is seen in the elastic cross section for $v_0=-240~\mathrm{MeV}\cdot\mathrm{fm}^{3}$.

\begin{figure}[ht]
\begin{center}
\includegraphics[width=80mm]{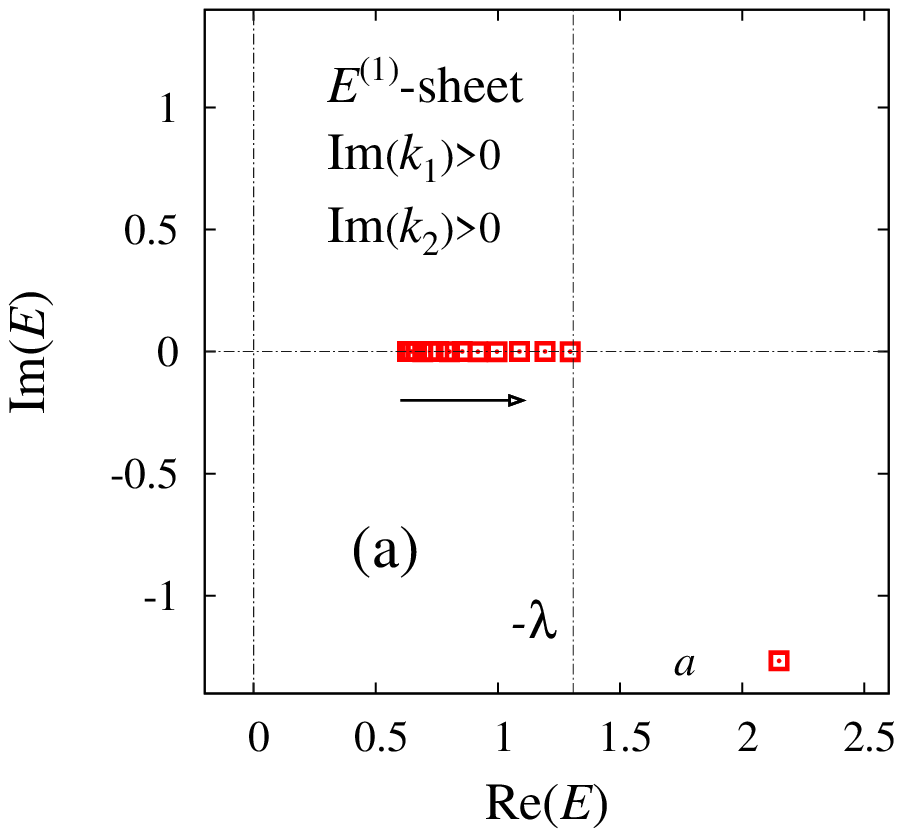}
\includegraphics[width=80mm]{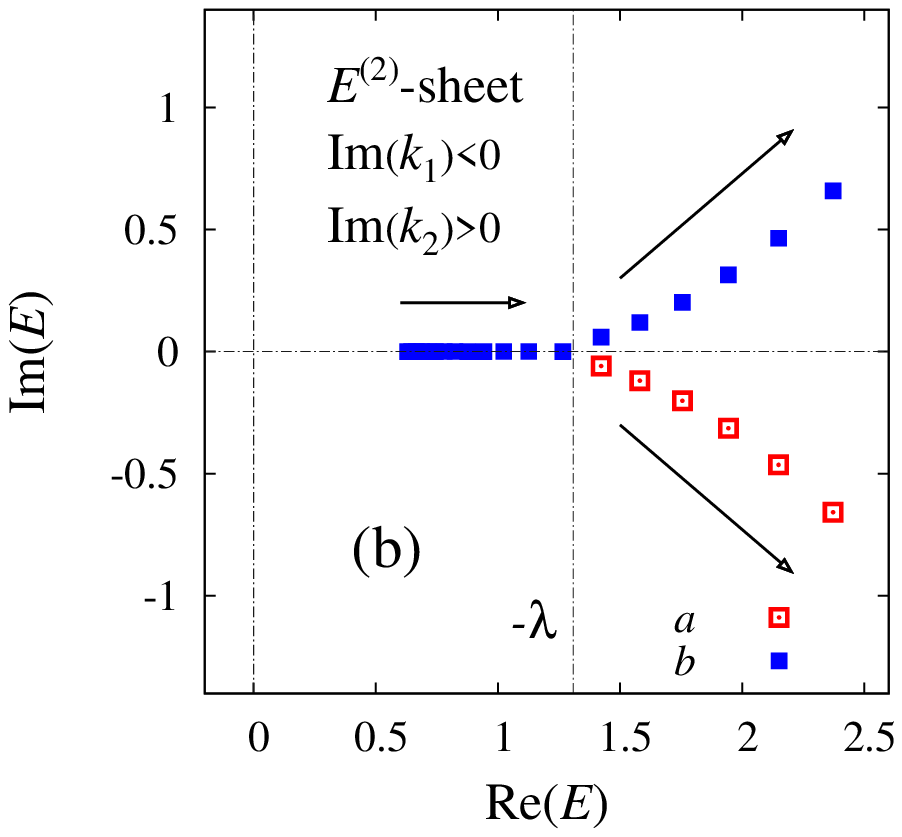}
\caption{Trajectory of the S-matrix poles in the scattering state ${}^{20}\mathrm{C}+n$ in the $s_{1/2}$-wave. (a) The  first Riemann sheet $E^{(1)}$ with $\mathrm{Im}\left(k_1 \right)>0$ and $\mathrm{Im}\left(k_2 \right)>0$. (b) The second sheet $E^{(1)}$ with $\mathrm{Im}\left(k_1 \right)<0$ and $\mathrm{Im}\left(k_2 \right)>0$. The square symbols represent the position of the poles for varied values of $v_0=0, -20, \cdots, -360, -380~\mathrm{MeV}\cdot\mathrm{fm}^{3}$. The threshold energy $E=-\lambda$ (for $v_0=-240~\mathrm{MeV}\cdot\mathrm{fm}^{3}$) is also shown.}
\end{center}
\end{figure}

\begin{figure}[ht]
\begin{center}
\includegraphics[width=80mm]{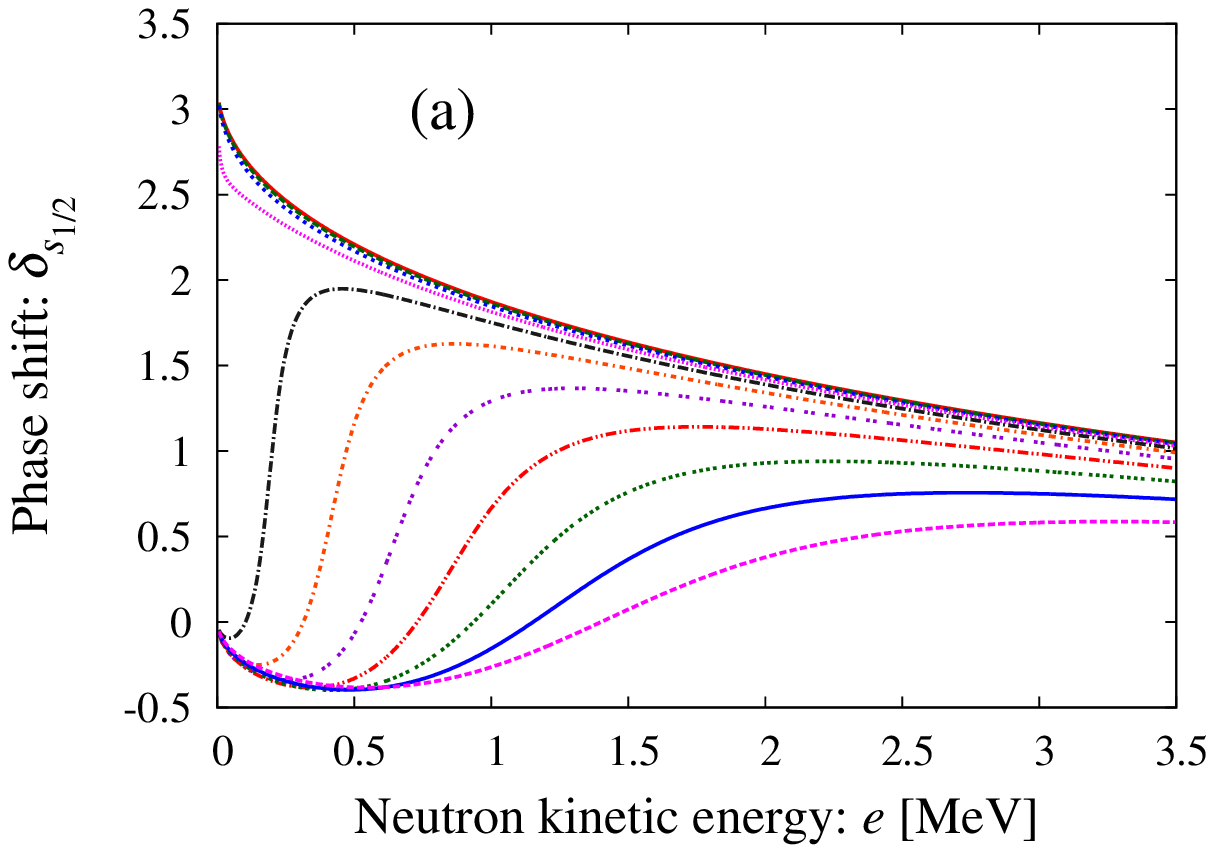}
\includegraphics[width=80mm]{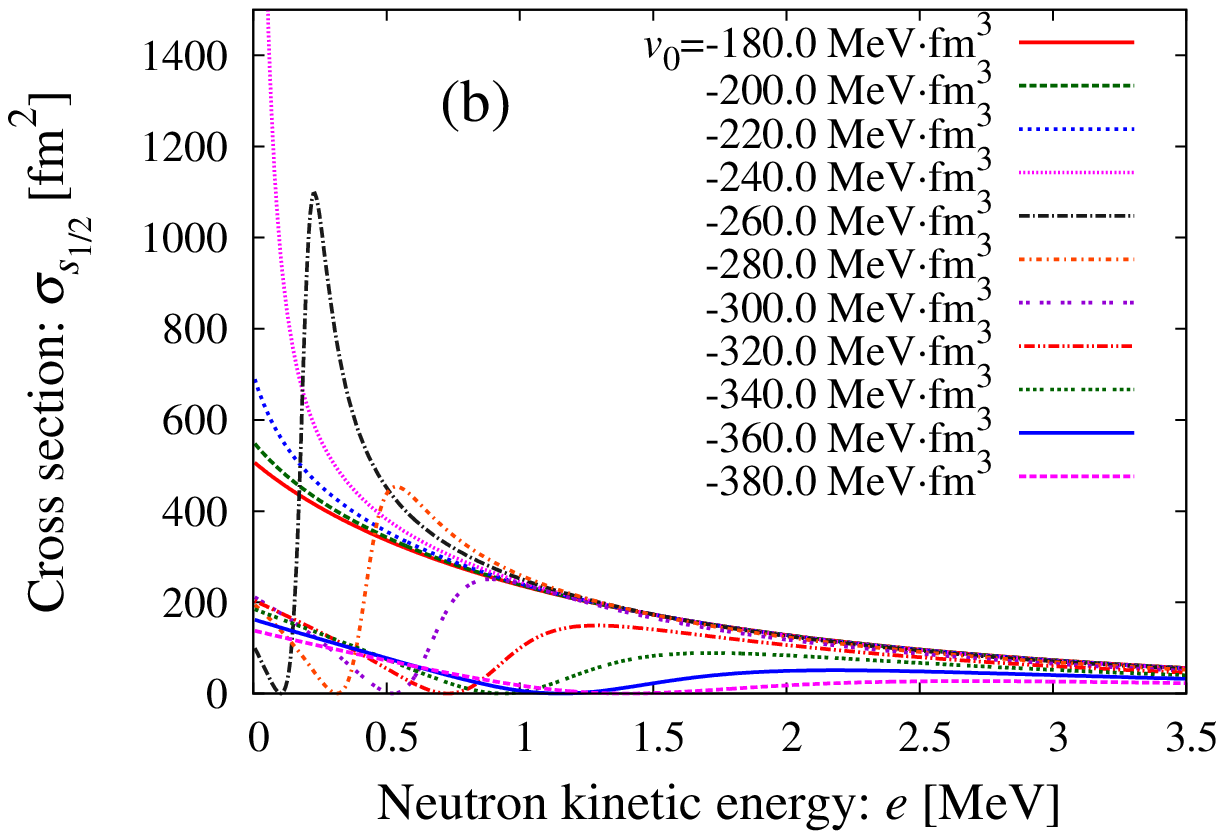}
\caption{Dependence on the pairing force strength $v_0$ of (a) the phase shift and (b) the elastic cross section for the scattering state ${}^{20}\mathrm{C}+n$ in the $s_{1/2}$-wave.}
\end{center}
\end{figure}

Figures 8 and 9 shows the results of the $d_{5/2}$ wave. We can see that there is $v_{0}$ dependence similar to Fig.~6. However, compared to the results of $s_{1/2}$-wave case, the poles in Fig.~8 are located near the real axis due to the centrifugal barrier. These poles make narrow peak structures in Fig.~9~(b) and Fig.~4. We note also that the quasiparticle energy of the $1d_{5/2}$ state depends more strongly on the pairing force strength $v_0$ than that of $2s_{1/2}$ as seen in Fig.~5~(a). It is because the effective pairing gap of the $1d_{5/2}$ is larger than that of $2s_{1/2}$. Consequently the resonance energy of the $1d_{5/2}$ is larger than that of the $2s_{1/2}$ at sizable pairing gap, e.g. at the representative value $v_0 =-320~\mathrm{MeV}\cdot\mathrm{fm}^{3}$ of the pairing strength.

\begin{figure}[ht]
\begin{center}
\includegraphics[width=80mm]{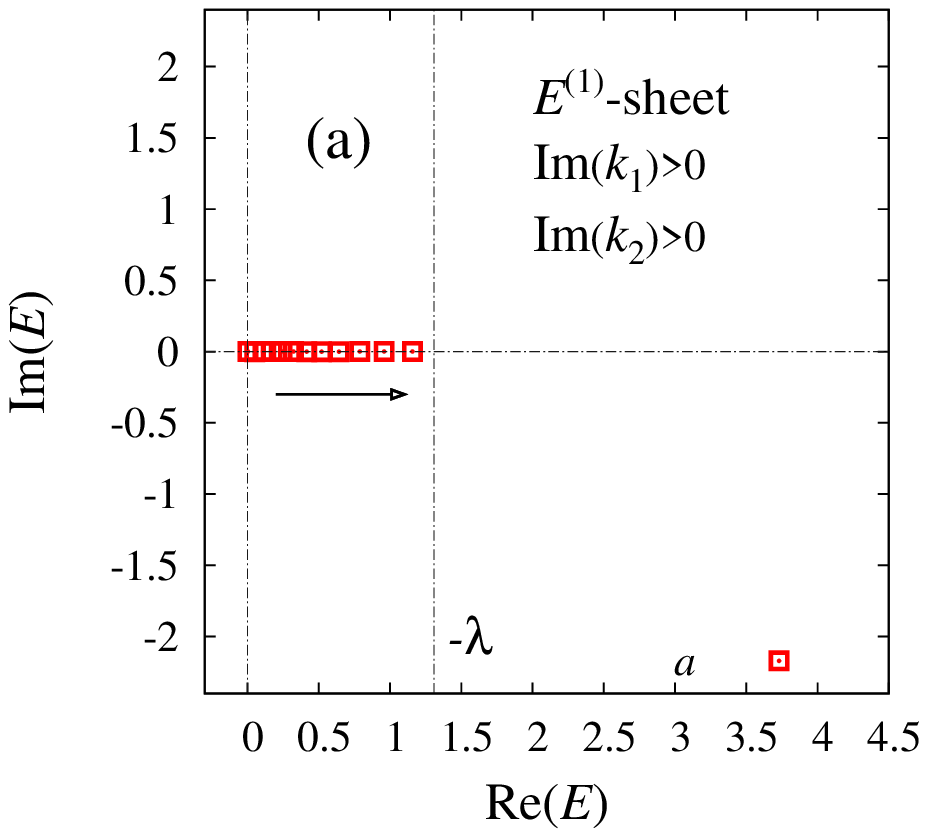}
\includegraphics[width=80mm]{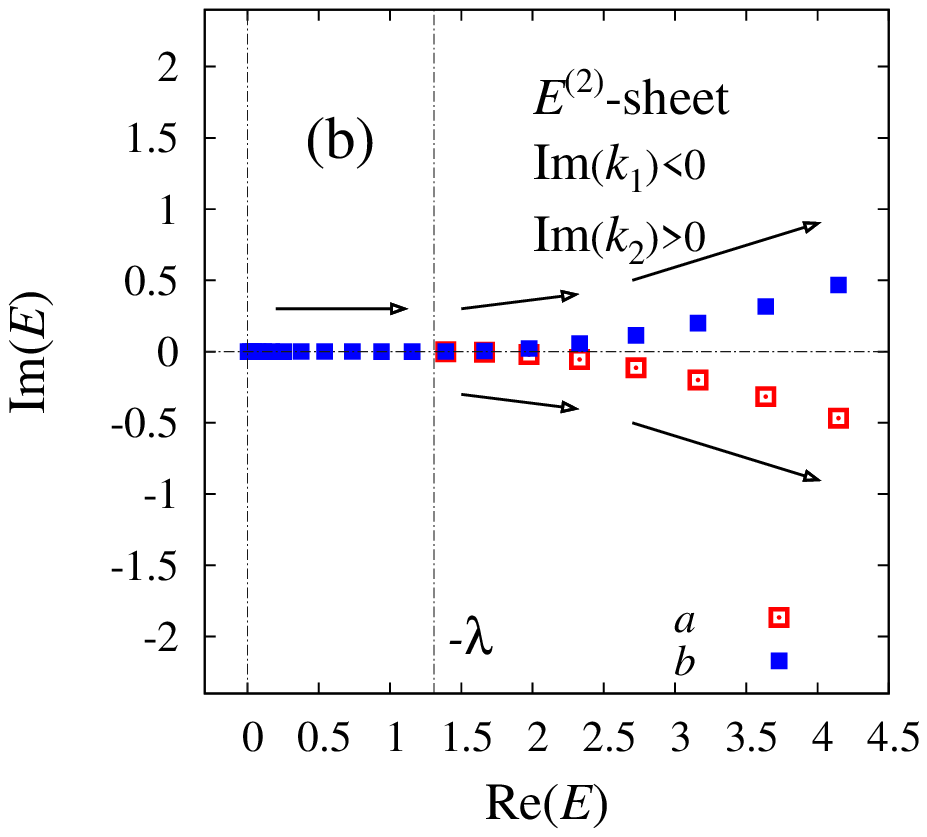}
\caption{The same as Fig.~6, but for the scattering state ${}^{20}\mathrm{C}+n$ in the $d_{5/2}$-wave.}
\end{center}
\end{figure}

\begin{figure}[ht]
\begin{center}
\includegraphics[width=80mm]{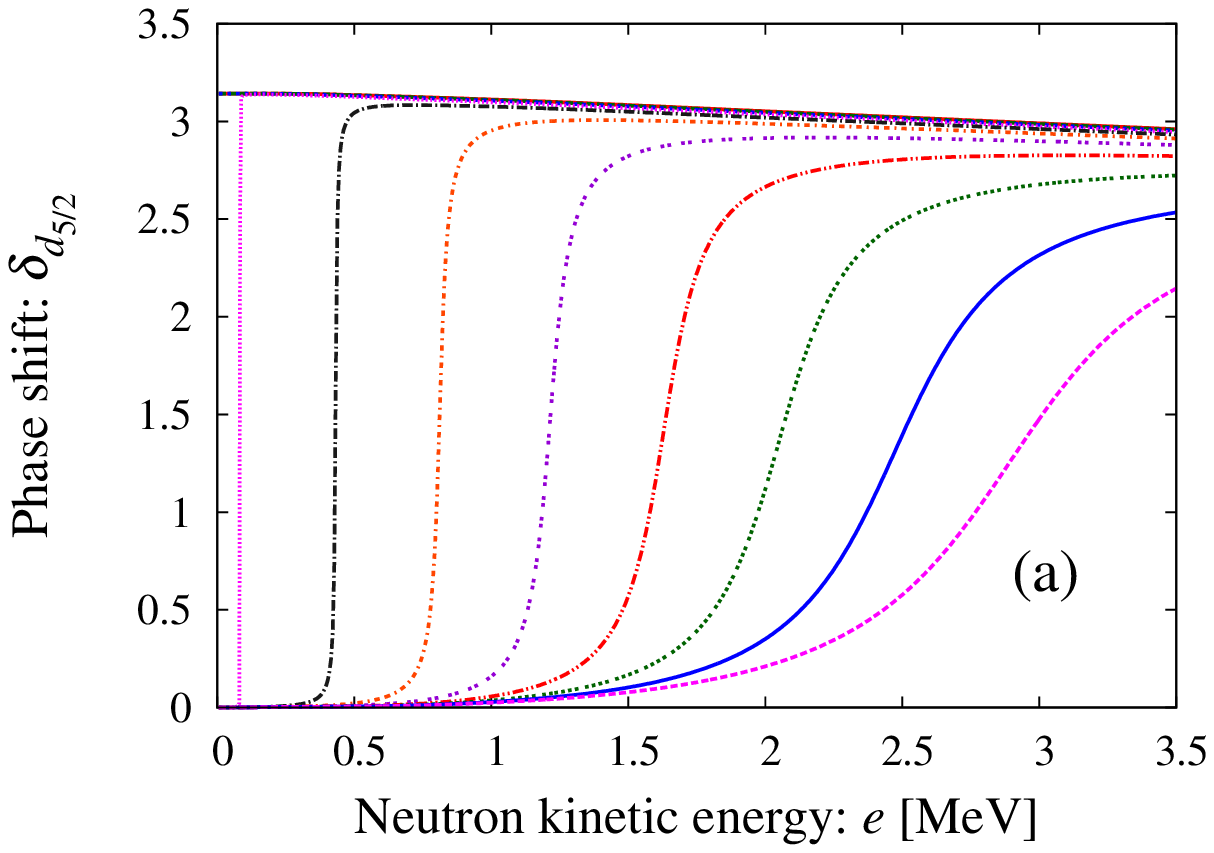}
\includegraphics[width=80mm]{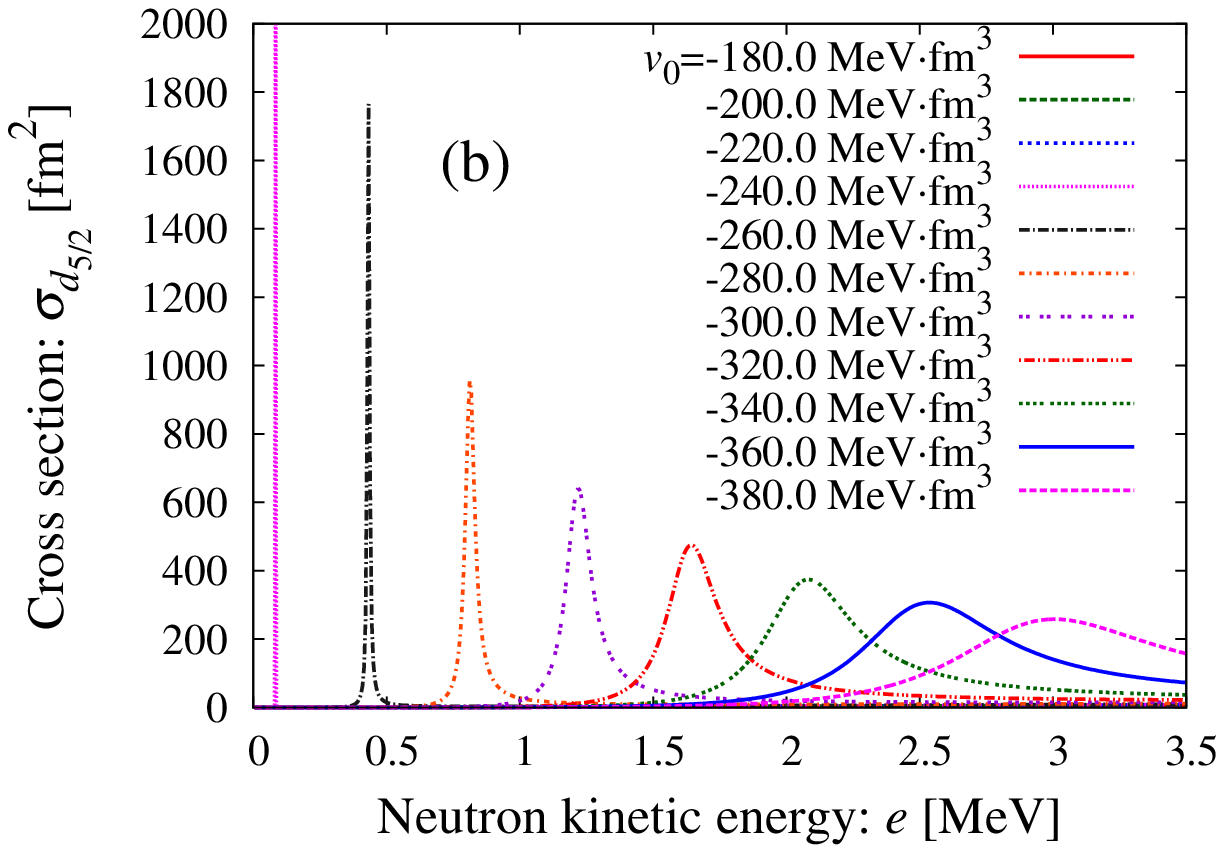}
\caption{The same as Fig.~7, but for the scattering state ${}^{20}\mathrm{C}+n$ in the $d_{5/2}$-wave.}
\end{center}
\end{figure}

\subsection{Decay spectrum of ${}^{21}\mathrm{C}$ in proton knockout reaction on ${}^{22}\mathrm{N}$}
Let us consider the case where ${}^{21}\mathrm{C}$ is produced via a proton knockout reaction on ${}^{22}\mathrm{N}$. Assuming that ${}^{22}\mathrm{N}$ is also influenced by the pairing correlation and is described in the mean-field framework of the HFB, the ground state configuration may be given as a two-quasiparticle configuration, one for neutron and the other for proton, since ${}^{22}\mathrm{N}$ is an odd-odd nucleus. Experimental spin parity $J^{\pi}$ of the ground state and the first excited state of ${}^{22}$N are $0^{-}$ and $1^{-}~(E_{1}=183\pm16~ \mathrm{keV})$, respectively~\cite{NNDC}, and it suggests that relevant quasiparticle states may be $\nu 2s_{1/2}$ and $\pi 1p_{1/2}$ states for neutrons and protons, respectively. We describe the ground state of ${}^{22}$N as a two-quasiparticle configuration $\nu 2s_{1/2} \otimes \pi 1p_{1/2}$: 
\begin{equation}
\left|\Psi_{{}^{22}\mathrm{N}}\right\rangle=
\left( \beta^{({}^{22}\mathrm{N})}_{\nu 2s_{1/2}} \right)^{\dagger}
\left( \beta^{({}^{22}\mathrm{N})}_{\pi 1p_{1/2}} \right)^{\dagger}
\left|\Phi_\nu^{({}^{22}\mathrm{N})}\right\rangle \otimes \left|\Phi_{\pi}^{({}^{22}\mathrm{N})}\right\rangle.
\end{equation}

The doorway state produced in the proton knockout reaction on ${}^{22}\mathrm{N}$ may be given by a configuration in which a proton is removed from the ground state of ${}^{22}\mathrm{N}$:
\begin{eqnarray}
&&|\Psi_{\mathrm{DWS}}\rangle \nonumber\\
&&=a_{1}'c^{({}^{22}\mathrm{N})}_{\pi 1p_{1/2}}\left|\Psi_{{}^{22}\mathrm{N}}\right\rangle+ a_{2}'c^{({}^{22}\mathrm{N})}_{\pi 1p_{3/2}}\left|\Psi_{{}^{22}\mathrm{N}}\right\rangle+\cdots \nonumber \\
&&=\left( \beta^{({}^{22}\mathrm{N})}_{\nu 2s_{1/2}} \right)^{\dagger}\left|\Phi_\nu^{{}^{22}\mathrm{N}}\right\rangle \otimes\left[
a_{1}'c^{({}^{22}\mathrm{N})}_{\pi 1p_{1/2}}\left( \beta^{({}^{22}\mathrm{N})}_{\pi 1p_{1/2}} \right)^{\dagger}\left|\Phi_{\pi}^{({}^{22}\mathrm{N})}\right\rangle\right.\nonumber \\
&& \hspace{10mm} +\left.a_{2}'c^{({}^{22}\mathrm{N})}_{\pi 1p_{3/2}}\left( \beta^{({}^{22}\mathrm{N})}_{\pi 1p_{1/2}} \right)^{\dagger}
\left|\Phi_{\pi}^{({}^{22}\mathrm{N})}\right\rangle+\cdots \right] .
\end{eqnarray}

The doorway state has finite overlap with scattering state of ${}^{20}\mathrm{C}+n$ with the $s$-wave neutron, and its overlap amplitude is given by
\begin{eqnarray}
&&\langle\Psi_{\mathrm{DWS}}|\Psi_{{}^{20}\mathrm{C}+n,s_{1/2}}(e)\rangle \nonumber\\
&&= S_1'\left\langle\Phi_{\nu}^{({}^{22}\mathrm{N})}\right|
\beta^{({}^{22}\mathrm{N})}_{\nu 2s_{1/2}}\
\left( \beta^{({}^{20}\mathrm{C})}_{\nu s_{1/2}}(E)\right)^{\dagger}\left|\Phi_{\nu}^{({}^{20}\mathrm{C})}\right\rangle,
\end{eqnarray}
with a constant
$S_1'=a_{1}'^*\left\langle\Phi_{\pi}^{{}^{22}\mathrm{N}}\right|\beta_{\pi 1p_{1/2}}^{({}^{22}\mathrm{N})}\left( c_{\pi 1p_{1/2}} ^{({}^{22}\mathrm{N})}\right)^{\dagger}\left|\Phi_{\pi}^{({}^{20}\mathrm{C})}\right\rangle$. Similarly to the neutron removal from ${}^{22}\mathrm{C}$, we evaluate the matrix element in Eq.~(25) using the diagonal approximation (Appendix A). Then we have 
\begin{eqnarray}
&&\langle\Psi_{\mathrm{DWS}}|\Psi_{{}^{20}\mathrm{C}+n,s_{1/2}}(e)\rangle \nonumber\\
&&\approx
S_1'\sum_{\sigma}\int d\bm{r}\left[ \left(\varphi^{({}^{22}\mathrm{N})}_{1,\nu 2s_{1/2}}(\bm{r}\sigma)\right)^{\ast}\varphi^{({}^{20}\mathrm{C})}_{1,\nu s_{1/2}}(\bm{r}\sigma,E)\right.\nonumber\\
&&\quad\left.+\left(\varphi^{({}^{22}\mathrm{N})}_{2,\nu 2s_{1/2}}(\bm{r}\sigma)\right)^{\ast}\varphi^{({}^{20}\mathrm{C})}_{2,\nu s_{1/2}}(\bm{r}\sigma,E) \right],
\end{eqnarray}
which is essentially an overlap integral of quasiparticle wave functions of the bound quasineutron state $2s_{1/2}$ in ${}^{22}\mathrm{N}$ and of the unbound quasineutron state with $s_{1/2}$ interacting with ${}^{20}\mathrm{C}$. 

Numerical calculation is performed in the following way. We first perform the HFB calculation for ${}^{22}\mathrm{N}$ to obtain the two-quasiparticle configuration, Eq.~(23). Here we have chosen the pairing force strength $v_{0}({}^{22}\mathrm{N})=-220.0~\mathrm{MeV}\cdot\mathrm{fm}^{3}$, which is smaller than that adopted or ${}^{20}\mathrm{C}$ and ${}^{22}\mathrm{C}$, in order to simulate the blocking effect caused by the quasiparticle excitation both in neutrons and protons. The result is shown in Table~III. The neutron average pairing gap is $\Delta \sim 0.9$ MeV, about one thirds of those in ${}^{20}\mathrm{C}$ and ${}^{22}\mathrm{C}$. The quasineutron state of the lowest energy is the $2s_{1/2}$ state with excitation energy $E_{2s_{1/2}}=1.18$ MeV, and the lowest quasiproton state is $1p_{1/2}$ with $E_{1p_{1/2}}=0.69$ MeV. This is consistent with our assumption that the ground state of ${}^{22}\mathrm{N}$ is $\nu 2s_{1/2} \otimes \pi 1p_{1/2}$. The energy difference between the quasineutron state and the threshold for unbound states (which corresponds to the neutron separation energy) is $|\lambda_n|-E_{2s_{1/2}}=1.34$ MeV, being qualitatively consistent with the experimental one-neutron separation energy $S_n=1.54 \pm 0.25$ MeV.

\begin{table}[ht]
\caption{Ground state properties of ${}^{22}\mathrm{N}$ obtained with the pairing force strength $v_0({}^{22}\mathrm{N})=-220~\mathrm{MeV}\cdot\mathrm{fm}^{3}$. $E_{2s_{1/2}}$ and $E_{1d_{5/2}}$ are the quasiparticle energy of the quasineutron states corresponding to the $2s_{1/2}$ and $1d_{5/2}$ orbits, respectively. See also the caption of Table.~I.}
\begin{ruledtabular}
\begin{tabular}{ccccc}
 & & Calc. & & Exp.\\ \hline
$e_{2s_{1/2}}$ [MeV] & & $-3.562$ && $-$\\
$e_{1d_{5/2}}$ [MeV] & & $-3.390$ && $-$\\
$E_{2s_{1/2}}$ [MeV] & & $1.183$ && $-$\\
$E_{1d_{5/2}}$ [MeV] & & $1.262$ && $-$\\
$\lambda$ [MeV] & & $-2.524$ && $-$\\
$S_{n}$ [MeV] && $1.340$ && $1.54\pm0.25$~\cite{NNDC}\\
$\sqrt{\left\langle r^{2}_{\mathrm{m}}\right\rangle}$ [fm] && $3.024$ && $3.08\pm0.12$~\cite{Bagchi2019}\\
$\Delta_{uv}$ [MeV] && $0.850$ && $-$\\
$\Delta_{uu}$ [MeV] && $0.933$ && $-$\\
\end{tabular}
\end{ruledtabular}
\end{table}

Figure~10 is the calculated decay spectrum for the $s_{1/2}$-wave decay of the doorway state produced by the proton removal from ${}^{22}\mathrm{N}$. The spectrum has a peak around 1 MeV, which originates from the quasiparticle resonance associated with neutron $2s_{1/2}$ orbit, i.e. the same resonance as those discussed for the decay spectrum in the neutron knockout reaction on ${}^{22}\mathrm{C}$. (The scattering state ${}^{20}\mathrm{C}+n$ is same as previous section which has the resonance parameters given in Table~I.) The profile of the peak is slightly different from that in the neutron knockout due to a difference in the overlap, Eq.~(13) v.s. Eq.~(26). A noticeable difference from the neutron knockout is that there is no components decaying to the $d_{5/2}$ and other partial waves. This is because the doorway state, Eq.~(24), produced by the proton knockout from the simple two-quasiparticle configuration, Eq.~(23), can have neutron scattering state ${}^{20}\mathrm{C}+n$ in the $s_{1/2}$-wave. The decay spectrum shown in Fig.~10 is qualitatively consistent with the experimental data~\cite{Leblond2015,Orr2016}.

\begin{figure}[ht]
\begin{center}
\includegraphics[width=80mm]{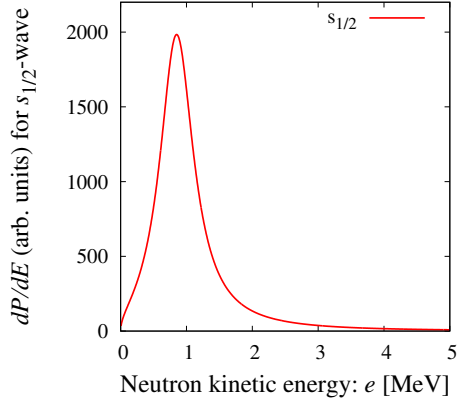}
\caption{Calculated decay spectrum $dP(e)/de$ in the $s_{1/2}$-wave for the ${}^{21}\mathrm{C}$ doorway state produced by the proton removal from ${}^{22}\mathrm{N}$.}
\end{center}
\end{figure}

We have treated the blocking effect in ${}^{22}\mathrm{N}$ in a simplified manner by reducing the pairing force strength $v_0$, which effectively reduces the pairing gap. This treatment may be checked by seeing the dependence on the parameter $v_0$. As shown in Fig.~11, the decay spectrum depends only weakly on the choice of $v_0$ as far as the $2s_{1/2}$ quasineutron state appears as a bound state with sizable binding energy $\gtrsim 0.5$ MeV with $|v_0|\lesssim260~\mathrm{MeV}\cdot\mathrm{fm}^{3}$ (see Fig.~12).

\begin{figure}[ht]
\begin{center}
\includegraphics[width=80mm]{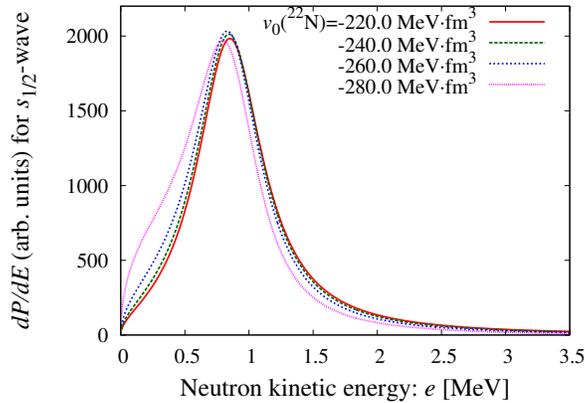}
\caption{The same as Fig.~10, but for the varied pairing force strength $v_0({}^{22}\mathrm{N})$.}
\end{center}
\end{figure}

\begin{figure}[ht]
\begin{center}
\includegraphics[width=80mm]{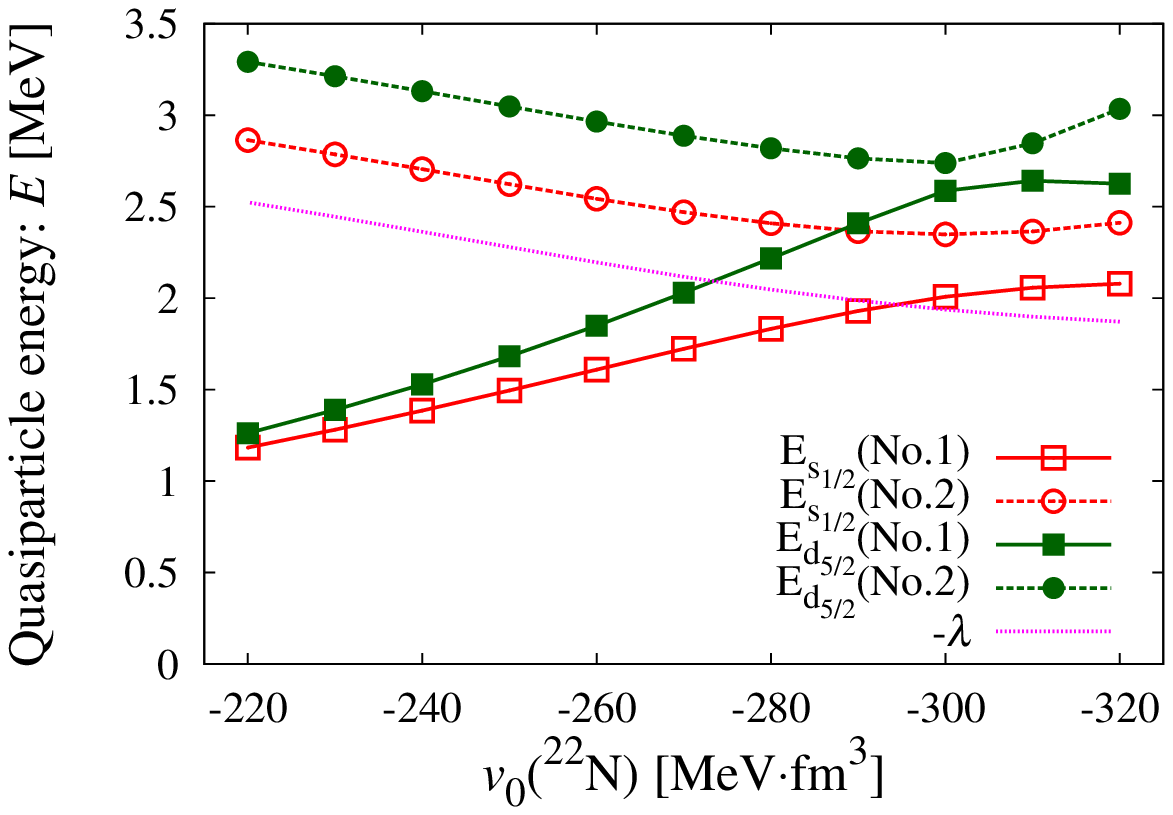}
\caption{
Dependence on the pairing force strength $v_0({}^{22}\mathrm{N})$ of the quasiparticle energies of the low-lying quasineutron states in the $s_{1/2}$- and $d_{5/2}$-waves in  ${}^{22}\mathrm{N}$. The threshold energy $-\lambda$ is also shown.}
\end{center}
\end{figure}

\section{Conclusions}
We have discussed the quasiparticle resonance appearing in the decay spectrum of unbound ${}^{21}\mathrm{C}$, which is produced in nucleon knockout reactions and then decays to ${}^{20}\mathrm{C}$ by emitting a neutron. Using the coordinate space HFB theory, we describe ${}^{21}\mathrm{C}$, a neighbor to the neutron drip-line, as an unbound one-quasineutron state built on the pair correlated ground state of ${}^{20}\mathrm{C}$. We then calculate the decay spectrum in terms of an overlap between the unbound one-quasiparticle state and a doorway state produced by the neutron (or proton) removal from ${}^{22}\mathrm{C}$ (or ${}^{22}\mathrm{N}$), all of which are described also in the same HFB framework.

We have shown that the quasiparticle resonances associated with the last bound neutron single-particle orbits, $2s_{1/2}$ and $1d_{5/2}$, show up as low-lying peaks in the decay spectrum of ${}^{21}\mathrm{C}$. A remarkable feature is that the quasiparticle resonance emerges in the decay channels of the $s_{1/2}$-wave neutron, i.e. in the $J^{\pi}=1/2^{+}$ state of ${}^{21}\mathrm{C}$. This is quite contrasting to the single-particle potential scattering of a neutron, which exhibits no $s$-wave resonance. The quasiparticle state under the influence of the pairing correlation has two components, each representing particle- and hole-like single-particle motion. Hence even the $s$-wave neutron can exhibit a resonance through a coupling to the hole component, which keeps a bound state character even if the quasiparticle itself represents a scattering neutron.
 
Another characteristic feature of the quasiparticle resonance is that the resonance energy and the resonance width depend sensitively on the strength of the pairing correlation since the coupling between the particle and hole components is governed by the pairing gap. In the present numerical analysis of ${}^{21}\mathrm{C}$ we find that a standard value of the pairing gap, approximately following the systematic value $\sim 12/\sqrt{A}$ MeV provides the $2s_{1/2}$ and $1d_{5/2}$ resonances, the resonance energy and width of which are consistent with the peaks found in the nucleon knockout reactions on ${}^{22}\mathrm{C}$ and ${}^{22}\mathrm{N}$. This suggests that the peaks found in the experiments may be realistic candidates of the quasiparticle resonance. Experimental confirmation of the quasiparticle resonances will provide information on the neutron pairing correlation in those nuclei. We note also the present HFB model describes qualitatively a clear difference between the experimental spectrum produced by the neutron knockout from ${}^{22}\mathrm{C}$ and that from the proton knockout from ${}^{22}\mathrm{N}$.
 
Finally we note that further investigations are required to confirm the quasiparticle resonance by comparing with the nucleon knockout experiments. First of all, the results in the present formulation does not describe the absolute value of the spectrum. We need to describe the cross section of the knockout process where the final state include ${}^{20}\mathrm{C}$ and a neutron with low relative energy.  This will allow us also to analyze, for example, the relative intensity populating the $2s_{1/2}$ and the $1d_{5/2}$ quasiparticle resonances in the reaction. Second, the phenomenological Woods-Saxon potential adopted in the present analysis needs to be improved, for example by replacing it with the selfconsistent mean-field. 

\begin{acknowledgments}
We thank Takashi Nakamura for fruitful discussions. This work is partially supported by Oita University President's Strategic Discretionary Fund. It is supported also by the JSPS KAKENHI, Grant No. 20K03945.
\end{acknowledgments}

\appendix

\section{Diagonal approximation for the overlap matrix elements}
Here we discuss the methods to evaluate the matrix elements which refer to two different HFB states. In the following we denote the two reference states $|\Phi_{0}\rangle$ and $|\Phi_{1}\rangle$. 

First we consider the matrix element $\langle\Phi_{1}|c^{(1)\dagger}_{i}\beta^{(0)\dagger}_{k}|\Phi_{0}\rangle$, where $c^{(1)\dagger}_{i}$ is a nucleon creation operator for single-particle state $i$ defined with respect to the mean-field associated with the reference state $|\Phi_{1}\rangle$ while $\beta^{(0)\dagger}_{k}$ is a creation operator of the quasiparticle state $k$ defined for the other reference state $|\Phi_{0}\rangle$. The nucleon operator and the quasiparticle operator can be expressed in terms of the field operators $(\psi^\dagger(x),\psi(x))$ and the single-particle wave function $\varphi_i^{(1)}$ and the quasiparticle wave function $(\varphi_{1,k}^{(0)}(x),\varphi_{2,k}^{(0)}(x))$ as
\begin{eqnarray}
c^{(1)\dagger}_{i}&&=\int dx\varphi^{(1)}_{i}(x)\psi^{\dagger}(x), \\
\beta^{(0)\dagger}_{k}&&=\int dx\left[ \varphi^{(0)}_{1,k}(x)\psi^{\dagger}(x)+\varphi^{(0)}_{2,k}(x)\psi(\tilde{x}) \right].
\end{eqnarray}
Here $x=\bm{r}\sigma$ is a collective notation of the coordinate spin variables. Note also $\displaystyle\int dx=\displaystyle\int d\bm{r} \sum_\sigma$ and $\psi(\tilde{x})=(-2\sigma)\psi(\bm{r} -\sigma)$.

The matrix element can be calculated as 
\begin{eqnarray}
&&\langle\Phi_{1}|c^{(1)\dagger}_{i}\beta^{(0)\dagger}_{k}|\Phi_{0}\rangle \nonumber\\
&&=\int dx \varphi^{(1)}_{i}(\tilde{x})\langle\Phi_{1}|\psi^{\dagger}(\tilde{x})\beta^{(0)\dagger}_{k}|\Phi_{0}\rangle\nonumber\\
&&= \int dx \varphi^{(1)}_{i}(\tilde{x})\sum_{k^{\prime}}
\langle\Phi_{1}|\psi^{\dagger}(\tilde{x})\beta^{(1)\dagger}_{k^{\prime}}|\Phi_{0}\rangle \left(U^{-1}\right)_{k^{\prime}k} \nonumber\\
&&=\langle\Phi_{1}|\Phi_{0}\rangle\int dx \varphi^{(1)}_{i}(\tilde{x})\sum_{k^{\prime}}\varphi^{(1)}_{2,k^{\prime}}(x)\left(U^{-1}\right)_{k^{\prime}k} .
\end{eqnarray}
Here $U_{k^{\prime}k}$ is matrix elements defining the generalized Bogoliubov transformation between  $\beta^{(0)\dagger}_{k}$ and
$\beta^{(1)\dagger}_{k^{\prime}}$, the quasiparticle operator referring to $|\Phi_{1}\rangle$:
\begin{equation}
\beta^{(1)\dagger}_{k}=\sum_{k^{\prime}}\left[ U_{k^{\prime}k}\beta^{(0)\dagger}_{k^{\prime}}+V_{k^{\prime}k}\beta^{(0)}_{k^{\prime}} \right],
\end{equation}
and it can be calculated as
\begin{eqnarray}
U_{k^{\prime}k}&=&\left\{ \beta^{(1)\dagger}_{k^{\prime}},~\beta^{(0)\dagger}_{k} \right\}\nonumber\\
&=&\int dx\varphi^{(0)\ast}_{1,k^{\prime}}(x)\varphi^{(1)}_{1,k}(x)+\int dx\varphi^{(0)\ast}_{2,k^{\prime}}(x)\varphi^{(1)}_{2,k}(x).\nonumber\\
\end{eqnarray}
The result, Eq.~(A3), is a variant of the Onishi formula~\cite{RingSchuck,Onishi1966}. Note also $\langle\Phi_{1}|\Phi_{0}\rangle =\sqrt{\mathrm{det}U}$. The above expressions can be calculated numerically with no approximation if the quasiparticle states $k$ and $k^{\prime}$ are discretized. However, it cannot be used in the case where the quasiparticle states have a continuum spectrum as the inverse matrix $U^{-1}$ is not calculated.

To overcome this problem we introduce an approximation
\begin{equation}
\langle\Phi_{1}|\psi^{\dagger}(\tilde{x})\beta^{(0)\dagger}_{k}|\Phi_{0}\rangle
\approx \varphi^{(0)}_{2,k}(x)
\end{equation}
with which the matrix element under discussion is given by
\begin{equation}
\langle\Phi_{1}|c^{(1)\dagger}_{i}\beta^{(0)\dagger}_{k}|\Phi_{0}\rangle\approx\int dx \varphi^{(1)}_{i}(\tilde{x})\varphi^{(0)}_{2,k}(x).
\end{equation}
This expression can be used both for discretized and continuum spectra. We have checked numerically the accuracy of this replacement, which we call the {\it diagonal approximation}. An example is shown in Fig.~13. 

\begin{figure}[ht]
\begin{center}
\includegraphics[width=80mm]{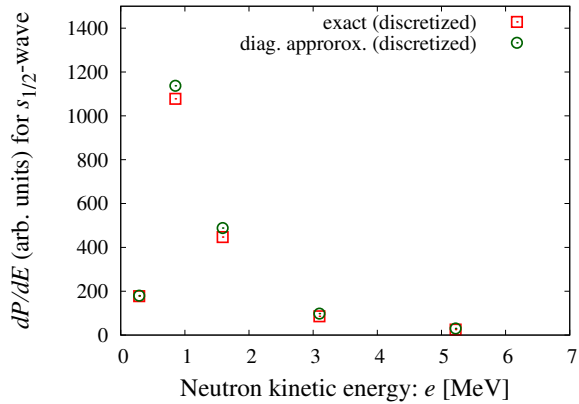}
\caption{The decay spectrum $dP(e)/de$ of the ${}^{21}\mathrm{C}$ doorway state for the $s_{1/2}$ wave, calculated with the same conditions as that in Fig.1, except that the unbound quasineutron states are discretized. The red square symbol is the calculation using the exact expression (A3) while the green circles are that with the diagonal approximation (A7). }
\end{center}
\end{figure}

The expression for the matrix element $\langle\Phi_{1}|\beta^{(1)}_{k^{\prime}}\beta^{(0)\dagger}_{k}|\Phi_{0}\rangle$ is obtained in the same way, and given by
\begin{widetext}
\begin{eqnarray}
&&\langle\Phi_{1}|\beta^{(1)}_{k^{\prime}}\beta^{(0)\dagger}_{k}|\Phi_{0}\rangle\nonumber\\
&&=\int dx \varphi^{(1)\ast}_{1,k^{\prime}}(x)\langle\Phi_{1}|\psi(x)\beta^{(0)\dagger}_{k}|\Phi_{0}\rangle+\int dx \varphi^{(1)\ast}_{2,k^{\prime}}(x)\langle\Phi_{1}|\psi^{\dagger}(\tilde{x})\beta^{(0)\dagger}_{k}|\Phi_{0}\rangle\nonumber\\
&&=\langle\Phi_{1}|\Phi_{0}\rangle \left[ \int dx \varphi^{(1)\ast}_{1,k^{\prime}}(x)\sum_{k^{\prime\prime}}\varphi^{(1)}_{1,k^{\prime\prime}}(x)\left(U^{-1}\right)_{k^{\prime\prime}k}+\int dx \varphi^{(1)\ast}_{2,k^{\prime}}(x)\sum_{k^{\prime\prime}}\varphi^{(1)}_{2,k^{\prime\prime}}(x)\left(U^{-1}\right)_{k^{\prime\prime}k}\right].
\end{eqnarray}
\end{widetext}
Applying the diagonal approximation~(A6) to Eq.~(A8), we obtain
\begin{eqnarray}
&&\langle\Phi_{1}|\beta^{(1)}_{k^{\prime}}\beta^{(0)\dagger}_{k}|\Phi_{0}\rangle\nonumber\\
&&\approx\int dx \varphi^{(1)\ast}_{1,k^{\prime}}(x)\varphi^{(0)}_{1,k}(x)+\int dx \varphi^{(1)\ast}_{2,k^{\prime}}(x)\varphi^{(0)}_{2,k}(x).\nonumber\\
\end{eqnarray}

\bibliography{manuscript_kobayashi_matsuo}

\end{document}